\documentclass[preprintnumbers,superscriptaddress,amsmath,amssymb,twocolumn]{revtex4-2}
\usepackage{float}
\usepackage{graphicx}
\usepackage{dcolumn}
\usepackage{bm}
\usepackage{epstopdf}
\usepackage{xcolor}
\usepackage{textcomp}
\usepackage{soul}
\usepackage{csquotes}
\usepackage{amsbsy}
\usepackage{amssymb}
\usepackage{mathrsfs} 
\usepackage{amsmath}
\usepackage{bigints} 
\usepackage{amsthm,amssymb} 
\usepackage[utf8]{inputenc}
\usepackage[shortlabels]{enumitem}
\usepackage{float}
\usepackage{mathrsfs,bigints,mathtools,dsfont}
\usepackage[colorlinks=true,linkcolor=blue,citecolor=blue]{hyperref}%
\usepackage[toc]{appendix}
\newcommand\norm[1]{\left\lVert#1\right\rVert}

\begin{document}
	
\title{{Global synchronization in generalized multilayer higher-order networks}}
\author{Palash Kumar Pal}\affiliation{Physics and Applied Mathematics Unit, Indian Statistical Institute, 203 B. T. Road, Kolkata 700108, India}
\author{Md Sayeed Anwar}\affiliation{Physics and Applied Mathematics Unit, Indian Statistical Institute, 203 B. T. Road, Kolkata 700108, India}
\author{Matja{\v z} Perc}
\affiliation{Faculty of Natural Sciences and Mathematics, University of Maribor,
Koro{\v s}ka cesta 160, 2000 Maribor, Slovenia}\affiliation{Community Healthcare Center Dr. Adolf Drolc Maribor, Vo{\v s}njakova ulica 2, 2000 Maribor, Slovenia}\affiliation{Complexity Science Hub Vienna, Josefst{\"a}dterstra{\ss}e 39, 1080 Vienna, Austria}\affiliation{Department of Physics, Kyung Hee University, 26 Kyungheedae-ro, Dongdaemun-gu, Seoul, Republic of Korea}
\author{Dibakar Ghosh}\email{dibakar@isical.ac.in}\affiliation{Physics and Applied Mathematics Unit, Indian Statistical Institute, 203 B. T. Road, Kolkata 700108, India}

\begin{abstract}
Networks incorporating higher-order interactions are increasingly recognized for their ability to introduce novel dynamics into various processes, including synchronization. Previous studies on synchronization within multilayer networks have often been limited to specific models, such as the Kuramoto model, or have focused solely on higher-order interactions within individual layers. Here, we present a comprehensive framework for investigating synchronization, particularly global synchronization, in multilayer networks with higher-order interactions. Our framework considers interactions beyond pairwise connections, both within and across layers. We demonstrate the existence of a stable global synchronous state, with a condition resembling the master stability function, contingent on the choice of coupling functions. Our theoretical findings are supported by simulations using Hindmarsh-Rose neuronal and R\"{o}ssler oscillators. These simulations illustrate how synchronization is facilitated by higher-order interactions, both within and across layers, highlighting the advantages over scenarios involving interactions within single layers.   
\end{abstract}	
\maketitle		
\section{Introduction}\label{sec1}
In recent decades, the study of complex networks has gained considerable traction, emerging as a prominent area of research. This surge in interest can be attributed to their remarkable capacity to model interconnected dynamical systems across various fields, including physics, biology, ecology, social sciences, and engineering \cite{strogatz2001exploring,boccaletti2006complex,wang2003complex}. These networks are comprised of nodes, representing entities or elements, and links, representing connections or pairwise interactions between them. Furthermore, researchers have introduced the concept of multilayer networks to extend the traditional network framework. Many real-world systems can effectively be conceptualized as multilayer networks \cite{kivela2014multilayer,boccaletti2014structure}. Examples include transportation networks \cite{cardillo2013modeling}, neuronal networks in the brain \cite{majhi2019chimera,rakshit2018synchronization}, and various types of social networks \cite{szell2010multirelational}. A multilayer network consists of individual networks, each with its set of nodes and links (referred to as intralayer links), interconnected through interlayer links. The representation of multilayer networks hinges on a fundamental assumption: the complex connections among individuals within and across layers are comprehensively elucidated through pairwise links.
\par 
While pairwise interactions, such as interlayer and intralayer links, are foundational and have yielded valuable insights, real-world systems often involve more intricate relationships. Indeed, systems in the real world, spanning human communications in social networks to neuronal interactions in the brain, can be accurately depicted through multilayer networks where interactions frequently occur among groups of three or more individuals simultaneously \cite{vasilyeva2021multilayer,huang2019higher,battiston2021physics,beyond_pairwise}. For example, in neuronal networks, neurons are interconnected through electrical and chemical synapses, giving rise to a multilayer structure \cite{anwar2022stability,rakshit2018synchronization}. Moreover, recent findings emphasize the existence of group interactions among individual neurons \cite{amari2003synchronous,ince2009presence,tlaie2019high,parastesh2022synchronization,anwar2023neuronal}. Similarly, the process of epidemic spreading among people involves groups of three or more interacting through virtual and physical layers \cite{fan2022epidemics,liu2023epidemic}. Traditional pairwise interactions fall short in capturing these group dynamics, necessitating the consideration of higher-order interactions \cite{beyond_pairwise,boccaletti2023structure}. These complexities are represented through simplicial complexes (hypergraphs), consisting of simplices (hyperedges) of varying dimensions \cite{hypergraph1,bianconi2021higher}, where a simplex (hyperedge) of order $d$ denotes a set of $(d+1)$ nodes. Recently, higher-order interactions have garnered significant attention from researchers due to their influence on collective phenomena \cite{zhang2023higher,anwar2024collective,skardal2020higher,alvarez2021evolutionary,anwar2024self,ghosh2023first} across various fields, including epidemiology \cite{wang2024epidemic,iacopini2019simplicial}, ecological systems \cite{grilli2017higher}, consensus dynamics \cite{neuhauser2021consensus}, and pattern formation \cite{muolo2023turing}.
\par  
One captivating collective phenomenon that has garnered significant attention in the realm of multilayer networks is synchronization. Synchronization refers to the remarkable ability of coupled individuals to self-organize and exhibit collective harmony in their behavior \cite{synchronization2,synchronization3}. Various types of synchronization phenomena have been identified within multilayer networks, including cluster synchronization \cite{pecora2014cluster,sorrentino2016complete}, antiphase synchronization \cite{chowdhury2021antiphase}, mixed synchronization \cite{pal2023mixed}, explosive synchronization \cite{zhang2015explosive}, intralayer synchronization \cite{anwar2022intralayer,anwar2022stability}, interlayer synchronization \cite{sevilla2016inter}, and relay interlayer synchronization \cite{PhysRevE.108.054208,anwar2021relay}. Each type of synchronization represents unique and intriguing behaviors exhibited by multilayer networks, making them an exciting and fertile area of exploration in network research. However, most prior works on synchronization in multilayer frameworks have considered interactions between individuals within and across layers to be pairwise, represented by links. Only a few have explored the impact of higher-order interactions on synchronization in multilayer networks \cite{anwar2022intralayer,anwar2022stability,skardal_multi,rathore2023synchronization,jalan2022multiple}. However, these studies have assumed that group interactions are confined to individuals within the layers of a multilayer network, or that the underlying node dynamics are governed by Kuramoto oscillators. Recently, authors in \cite{skardal_multi} investigated a multilayer framework where interactions both within and across layers are considered beyond pairwise. However, in this study, the authors assumed all-to-all connection mechanisms among individuals and focused solely on the case of the Kuramoto oscillator. Thus, there remains ample opportunity to explore synchronization phenomena in generalized higher-order multilayered systems beyond the conventional Kuramoto model and all-to-all framework.
\par 
In this study, we aim to address this gap by exploring the phenomenon of synchronization, particularly global synchronization, within a more generalized multilayer higher-order network. This network encompasses interactions beyond pairwise connections, both within and across layers. Our proposed mathematical framework surpasses the constraints of all-to-all coupling configurations and specific oscillator models. Instead, we employ a collection of identical dynamical systems distributed across different layers, engaging in interactions within and across layers in groups of two or three (up to order three for simplicity). In this broad context, we demonstrate the emergence of global synchronization within the multilayer network, contingent upon the cancellation of coupling functions. We examine these coupling functions as either linear or nonlinear diffusive couplings. Subsequently, we establish the necessary condition for the stability of the synchronization solution, resulting in a system of coupled linear equations known as the master stability equation. Notably, our stability condition bears a resemblance to the well-established master stability function (MSF) form under specific circumstances, enabling the derivation of fully decoupled master stability equations. Finally, we validate our findings using random higher-order multilayer networks comprising paradigmatic coupled chaotic systems, such as the Hindmarsh-Rose neuronal models and chaotic R\"{o}ssler oscillators.
\par 
The paper is structured as follows: Section \ref{sec2} introduces our mathematical model for describing the generalized higher-order multilayer network. The linear stability analysis of the global synchronization state is presented in Section \ref{stability}. Section \ref{result} presents the numerical results corresponding to the two coupled systems, viz, Hindmarsh-Rose neurons and R\"{o}ssler oscillators. Finally, Section \ref{sec5} provides a summary and conclusion.

\section{Mathematical model of the generalized multilayer higher-order network}\label{sec2}
\begin{figure}[hpt]
    \centering{
    \includegraphics[scale=0.35]{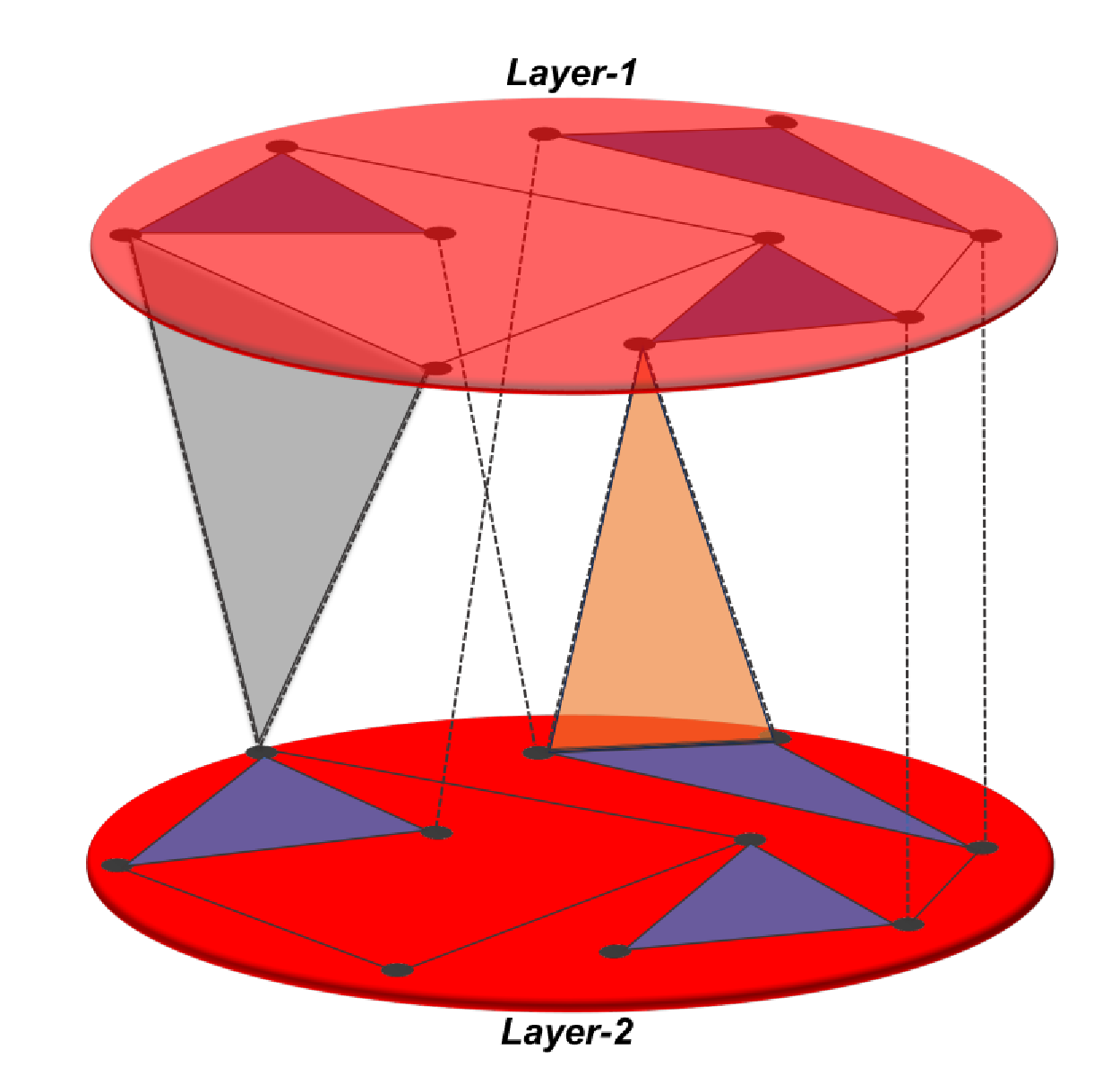}}
    \caption{{\bf Schematic diagram of the multilayer higher-order network}. There are two layers, Layer-$1$ and Layer-$2$, colored in red, each composed of $10$ nodes. The black solid circles denote the nodes. The solid black lines represent the intralayer pairwise interactions, and the dashed black lines represent the interlayer pairwise interactions. Triangles filled in blue represent the intralayer three-body interactions. The triangles filled in grey and orange represent interlayer three-body interactions. In the grey triangle, one node from Layer-$2$ is connected with two nodes from Layer-$1$. Whereas the orange triangle indicates the three-body interactions in which one node from Layer-$1$ participates with two nodes from Layer-$2$.}
    \label{fig1}
\end{figure}

We start by considering a $M$-layered higher-order multilayer network, which is schematized for two layers ($ M=2$) in Fig. \ref{fig1}. Each layer comprises $N$ number of nodes, and the $i$th node in one layer is identical with the $i$-node in all the other layers. Here, for simplicity's sake, we consider an identical number of nodes in each layer. The state of the $i$-th node in the $l$-th layer at time $t$ is given by the $d$-dimensional state vector ${\mathbf{X}}_{li}(t)$, where $l=1,2,\cdots, M$, and $i=1, 2, \cdots, N$. Now, we consider that the nodes in each layer are connected to one another through pairwise links ($1$-simplices) and groups of three ($2$-simplices) within and across the layers. 
The following coupled differential equations give the equation of motion governing the dynamics of the multilayer network:
\begin{multline}
\dot{{\bf{X}}}_{li}=f({\bf{X}}_{li})+\sum\limits_{l_1=1}^{M}\epsilon_1^{ll_1}\sum\limits_{j=1}^N {{A}}^{ll_1}_{ij}h^{(1)}({\bf{X}}_{l_1j},{\bf{X}}_{li})+\\ \sum\limits_{l_1=1}^M\sum\limits_{l_2=1}^M\epsilon_2^{ll_1l_2}\sum\limits_{j=1}^N\sum\limits_{k=1}^N{{B}}^{ll_1l_2}_{ijk}h^{(2)}({{\bf{X}}_{l_1j},{\bf{X}}_{l_2k},{\bf{X}}_{li}}),\\
l=1,2,\dots, M,\; \mbox{and}\; i=1,2,\dots, N.
    \label{equ1}
    \end{multline}
Here, each node's dynamics are identical and illustrated as $f:\mathscr{R}^d \rightarrow \mathscr{R}^d$. $\epsilon_{1}^{ll_{1}}$ is the pairwise coupling strength between the nodes of layers $l$ and $l_{1}$ connected through pairwise links. $\epsilon_{2}^{ll_{1}l_{2}}$ denotes the three-body coupling strength among the nodes of layers $l,l_{1},$ and $l_{2}$ interconnected via $2$-simplices. Certainly, when $l=l_{1}$, the pairwise interactions occur among the nodes of the same layer (i.e., intralayer connections); conversely, for $l \neq l_{1}$, they take place across two different layers (i.e., interlayer connections). Similarly, in the context of three-body interactions, when $l=l_{1}=l_{2}$, the triadic interactions happen among the nodes of a specific layer $l$ (i.e., intralayer three-body interactions). For $l=l_{1} \neq l_{2}$, two of the nodes in a group of three are from one layer, and the third one is from another layer. Whereas in the case of $l \neq l_{1} \neq l_{2}$, all the three nodes participating in a three-body interaction are from three different layers. The last two instances correspond to interlayer three-body interactions. For the sake of simplicity, we consider the pairwise coupling strengths as $\epsilon_{1}^{ll_{1}}=\epsilon_{1}$ if the pairwise interaction occurs within a specific layer, and $\epsilon_{1}^{ll_{1}}= \alpha \epsilon_{1}$ for interlayer connections. The same convention has been taken for the three-body coupling strengths, i.e., $\epsilon_{2}^{ll_{1}l_{2}}=\epsilon_{2}$ if the three-body interactions take place within a specific layer and $\epsilon_{2}^{ll_{1}l_{2}}=\alpha \epsilon_{2}$ when the three-body interactions are interlayer. Here, $\alpha$ is an arbitrary constant satisfying $0<\alpha<1$, indicating that for both pairwise and three-body interactions, the intralayer coupling strengths are more effective than the interlayer ones. Thus, the parameter $\alpha$ quantifies the strength of interlayer interactions in our study. Indeed, the described convention mirrors real-world scenarios where individuals within the same group tend to form stronger bonds among themselves compared to their interactions with individuals from different groups. ${{A}}^{ll_{1}}$ is a $N\times N$ matrix that describes how the nodes of the layers $l$ and $l_{1}$ are connected to one another through pairwise links. The elements ${{A}}^{l_1l_2}_{ij}=1$ if the nodes $i,j$ in the layers $l_1,l_2$, respectively are connected, and otherwise ${{A}}^{l_1l_2}_{ij}=0$. Thus, when $l=l_{1}$, it corresponds to the intralayer adjacency matrix, and for $l \neq l_{1}$ corresponds to the interlayer adjacency matrix. Similarly, ${{B}}^{ll_{1}l_{2}}$ is the adjacency tensor that provides information about which nodes participate in three-body interactions. The elements ${{B}}^{l_1l_2l_3}_{ijk}=1$ if there is a $2$-simplex interaction between the nodes $i,j,k$ in the layers $l_1,l_2$, and $l_3$, respectively, while ${{B}}^{l_1l_2l_3}_{ijk}=0$ if there are no such three-body interactions. $h^{(1)}:\mathscr{R}^d\times\mathscr{R}^d\rightarrow\mathscr{R}^d$, and $h^{(2)}:\mathscr{R}^d\times\mathscr{R}^d\times\mathscr{R}^d\rightarrow\mathscr{R}^d$ are the pairwise and three-body coupling functions such that $h^{(1)}({\bf x},{\bf x})=0$ and $h^{(2)}({\bf x},{\bf x},{\bf x})=0$, i.e., the coupling functions cancel out when the state of all connected individuals are equal. Thus, here we consider the pairwise and three-body coupling functions to be generalized diffusive, i.e., $h^{(1)}({\bf{X}}_{l_1i},{\bf{X}}_{l_2j})=H({\bf{X}}_{l_1i})-H({\bf{X}}_{l_2j})$ and $h^{(2)}({\bf{X}}_{l_1k},{\bf{X}}_{l_2j}, {{\bf{X}}_{l_3i}})=G({\bf{X}}_{l_1k},{\bf{X}}_{l_2j})-G({\bf{X}}_{l_3i},{\bf{X}}_{l_3i})$ \cite{gambuzza2021stability,anwar2024global}.
 
 \section{Stability analysis of the global synchronization state}\label{stability}
Here, we investigate the stability of the global synchronization state, wherein the nodes of the multilayer structure oscillate in unison. In mathematical terms, $\mathbf{X}_{li}=\mathbf{X}_{l^{'}j}$ for all $l,l'=1,2,\dots,M$ and $i,j=1,2,\dots,N$. The selection of diffusive pairwise and triadic coupling functions indeed ensures the existence and invariance of the global synchronization state in our multilayer system. 
\par Now, before proceeding to the stability analysis, it is beneficial to simplify the model. To do so, we set $\mathbf{X}_{li}= {\mathbf{X}_{i+(l-1)N}}$ where $(l=1,2,\dots, M)$, and $(i=1,2,\dots, N)$, effectively transforming the $M$ layered multilayer network with $N$ nodes in each layer into a network with $MN$ nodes. Thus, the model Eq.~\eqref{equ1} eventually becomes,
\begin{multline}
 \label{equ2}
\dot{\mathbf{X}}_i=f(\mathbf{X}_i)+\epsilon_1\sum\limits_{j=1}^{MN}\mathscr{A}_{ij}^{(1)}(H(\mathbf{X}_j)-H(\mathbf{X}_i)) \\
+\epsilon_2\sum\limits_{k=1}^{MN}\sum\limits_{j=1}^{MN}\mathscr{A}_{ijk}^{(2)}(G(\mathbf{X}_{j},\mathbf{X}_{k})-G(\mathbf{X}_{i},\mathbf{X}_{i})),\\ \;\;i=1,2,\dots,MN, 
\end{multline}
where $\mathscr{A}^{(1)}\in\mathscr{R}^{MN\times MN}$ is the block matrix given by,
{\begin{equation*}
\begin{bmatrix}
        {A}^{11} & \dots & {A}^{1M} \\
        \vdots & \ddots & \vdots \\
        {A}^{M1} & \dots & {A}^{MM}\\
\end{bmatrix},
\end{equation*}}
${A}^{l_1l_2}\in\mathscr{R}^{N\times N}$ such that the elements for $l_1,l_2=1,2,\dots,M$ and $i,j=1,2,\dots,N$ are defined as
\begin{equation*}
    \begin{array}{l}
        {{A}}^{l_1l_2}_{ij} = \begin{cases}
    1,~\text{if}~i\leftrightarrow_1j\\
    \alpha,~\text{if}~i\leftrightarrow_2j\\
    0,~\text{if}~i\nleftrightarrow_1j~\text{and}~i\nleftrightarrow_2 j.
\end{cases} 
    \end{array}
\end{equation*}
\par Here $i\leftrightarrow_1j$ means an intralayer link (i.e., $l_{1}=l_{2}$) exists between the nodes $i$ and $j$, and $i\nleftrightarrow_1j$ means no such link exists. $i\leftrightarrow_2j$ means there exists an interlayer link (i.e., $l_{1}\neq l_{2}$) between the nodes $i$ and $j$, and if there is no such link, then $i\nleftrightarrow_2j$. Similarly, the tensor $\mathscr{A}^{(2)}\in\mathscr{R}^{MN\times MN\times MN}$ is given by, 
\begin{equation*}
    \begin{array}{l}
        \mathscr{A}^{(2)}_{ijk} = \begin{cases}
    1~\text{if}~i,j,k\in \triangle_1,\\
    \alpha~\text{if}~i,j,k\in \triangle_2,\\
    0~\text{if}~i,j,k\notin \triangle_1,\triangle_2.
\end{cases} 
    \end{array}
\end{equation*}
The symbols $\triangle_1$ and $\triangle_2$ indicate an intralayer triangle {$(l_{1}=l_{2})$} and an interlayer triangle {$(l_{1}\neq l_{2})$}, respectively. If there are such triangles between the nodes $i,j,k$, then $i,j,k\in \triangle_1$ or, $i,j,k\in \triangle_2$, and if there are no such triangles, then $i,j,k\notin \triangle_1$ and $i,j,k\notin\triangle_2$.
\par Now, suppose $\mathbf{X}^s$ is the global synchronization state of the multilayer network. To analyze the stability of the given state, we consider small deviations around the synchronized state, denoted as $\mathbf{X}_{i}=\mathbf{X}^s+\delta\mathbf{X}_{i}$. Subsequently, we linearize Eq.~\eqref{equ2} with respect to these perturbations, which yields the variational equations as,
\begin{multline}        \delta\dot{\mathbf{X}}_{i}=\mathbf{J}f(\mathbf{X}^s)\delta\mathbf{X}_i+\epsilon_1\sum\limits_{j=1}^{MN}\mathscr{A}^{(1)}_{ij}\mathbf{J}H(\mathbf{X}^s)(\delta\mathbf{X}_{j}-\delta\mathbf{X}_{i})\\
+\epsilon_2\sum\limits_{j=1}^{MN}\sum\limits_{k=1}^{MN}\mathscr{A}^{(2)}_{ijk}\big[\mathbf{J}_1G(\mathbf{X}^s,\mathbf{X}^s)(\delta\mathbf{X}_{j}-\delta\mathbf{X}_{i}) \\
+\mathbf{J}_2G(\mathbf{X}^s,\mathbf{X}^s)(\delta\mathbf{X}_{k}-\delta\mathbf{X}_{i})\big],\\~~ i=1,2,\cdots,MN,
    \label{equ3}
\end{multline}
where $\mathbf{J}f(\mathbf{X}^s)$ and $\mathbf{J}H(\mathbf{X}^s)$ are the Jacobian matrices of the functions $f$ and $H$ evaluated at the synchronous solution. $\mathbf{J}_1G(\mathbf{X}^s,\mathbf{X}^s)$ and $\mathbf{J}_2G(\mathbf{X}^s,\mathbf{X}^s)$ are the Jacobians with respect to the first and second variables of the coupling function $G$ at the synchronization state. To proceed further, we use the zero-row sum Laplacian matrices corresponding to the pairwise and higher-order interactions \cite{gambuzza2021stability}. If  $\mathscr{L}^1$ and $\mathscr{L}^2$ are the Laplacians corresponding to the pairwise and three-body interactions, then we have 
\begin{equation*}
    \mathscr{L}^{(1)}_{ij} = \begin{cases}
    -\mathscr{A}^{(1)}_{ij} & \text{if } i \neq j, \\
    \sum\limits_{j=1}^{MN}\mathscr{A}^{(1)}_{ij} & \text{if } i = j,
\end{cases}
\end{equation*}
and
\begin{equation*}
\mathscr{L}^{(2)}_{ij} = \begin{cases}
    -\sum\limits_{k=1}^{MN}\mathscr{A}^{(2)}_{ijk} & \text{if } i \neq j, \\
    \sum\limits_{j=1}^{MN}\sum\limits_{k=1}^{MN}\mathscr{A}^{(2)}_{ijk} & \text{if } i = j.
\end{cases}    
\end{equation*}
Incorporating the definitions of these Laplacian matrices, the variational equation \eqref{equ3} can be rewritten as follows:
\begin{multline}
\delta\Dot{\mathbf{X}}_{i}=\mathbf{J}f(\mathbf{X}^{s})\delta\mathbf{X}_i-\sum\limits_{j=1}^{MN}\big[\epsilon_1\mathbf{J}H(\mathbf{X}^s)\mathscr{L}^{(1)}_{ij}+\\\epsilon_2(\mathbf{J}_1G(\mathbf{X}^s,\mathbf{X}^s)+\mathbf{J}_2G(\mathbf{X}^s,\mathbf{X}^s))\mathscr{L}^{(2)}_{ij}\big]\delta\mathbf{X}_j \\i=1,2,\cdots, MN.
    \label{equ4}
\end{multline}
Now considering the state variable $\delta\mathbf{X}=(\delta\mathbf{X}_1^T~\delta\mathbf{X}_2^T\dots\delta\mathbf{X}_{MN}^T)^T$ and using matrix Kronecker product $\otimes$, the variational equation \eqref{equ4} can be written in vectorial form  as
\begin{multline}
      \label{equ5}
        \delta\Dot{\mathbf{X}}=\mathbf{I}_{MN}\otimes\mathbf{J}f(\mathbf{X}^{s})\delta\mathbf{X}-\big[\epsilon_1\mathscr{L}^{(1)}\otimes \mathbf{J}H(\mathbf{X}^s) \\\\
        +\epsilon_2\mathscr{L}^{(2)}\otimes (\mathbf{J}_1G(\mathbf{X}^s,\mathbf{X}^s)+\mathbf{J}_2G(\mathbf{X}^s,\mathbf{X}^s)) \big] \delta\mathbf{X},
\end{multline}
where $()^T$ represents vector transpose and $\mathbf{I}$ is the identity matrix.
\par The coupled linearized variational equation \eqref{equ5} compromises two components: one governing motion along the synchronization manifold, referred to as parallel modes, and the other handling motion across the manifold, known as transverse modes. The stability of the synchronization state is contingent upon the convergence of all transverse modes to zero over time. To perform linear stability analysis, we thus need to decompose the variational equation into parallel and transverse modes and derive the condition for the latter's extinction. 
\par Now both the Laplacian matrices $\mathscr{L}^{(1)}$ and $\mathscr{L}^{(2)}$ are real symmetric matrices with zero row-sum. As a result, these matrices are diagonalizable by their basis of eigenvectors, possessing nonnegative real eigenvalues. Their common smallest eigenvalue is denoted as $\lambda_{1}=0$, and the corresponding eigenvectors form an orthonormal basis.
\par Therefore, to distinguish transverse and parallel modes in Eq. \eqref{equ5}, we project the perturbed variable $\delta \mathbf{X}$ onto the basis of eigenvectors $\mathbf{V}_{(1)}=[\mathbf{v}_1~\mathbf{v}_2\dots\mathbf{v}_{MN}]$, associated with the pairwise Laplacian $\mathscr{L}^{(1)}$, where $\mathbf{v}_{1}=(1/\sqrt{MN},1/\sqrt{MN},\dots,1/\sqrt{MN})^{T}$ is the eigenvector corresponding to the smallest eigenvalue $\lambda_{1}=0$. This projection is accomplished by introducing a new variable $\eta=(\mathbf{V}_{(1)}^{-1}\otimes\mathbf{I}_{d})\delta\mathbf{X}$ where $\eta=(\eta_1^{T}~\eta_2^{T}\dots\eta_{MN}^{T})^T$. The choice of the eigenvector basis is arbitrary. An alternative basis can be selected, and other sets of eigenvectors can be transformed into the chosen basis through a unitary matrix transformation. Thus, expressed in the introduced variables, the variational Eq. \eqref{equ5} undergoes the following transformation:
\begin{multline}
\label{eta_var_eq}
 \dot{\eta}= \mathbf{I}_{MN}\otimes\mathbf{J}f(\mathbf{X}^{s})\eta-\big[\epsilon_1 \Lambda\otimes \mathbf{J}H(\mathbf{X}^s) \\\\
        +\epsilon_2\Tilde{\mathscr{L}}^{(2)}\otimes (\mathbf{J}_1G(\mathbf{X}^s,\mathbf{X}^s)+\mathbf{J}_2G(\mathbf{X}^s,\mathbf{X}^s)) \big] \eta,
\end{multline}
where $\mathbf{V}_{(1)}^{-1}\mathscr{L}^{(1)}\mathbf{V}_{(1)}=\Lambda=diag\{0=\lambda_{1},\lambda_{2},\dots,\lambda_{MN}\}$ and $\Tilde{\mathscr{L}}^{(2)}=\mathbf{V}_{(1)}^{-1}\mathscr{L}^{(2)}\mathbf{V}_{(1)}$. Now, since the Laplacian $\mathscr{L}^{(2)}$ is a zero-row sum matrix, $\Tilde{\mathscr{L}}^{(2)}$ must have null first row and column.
The transformed variational Eq.~\eqref{eta_var_eq} then can be partitioned into two components as follows:
\begin{align}\label{eta_para}
    \dot{\eta}_{1}= \mathbf{J}f(\mathbf{X}^{s}) \eta_{1},
\end{align}
and
\begin{multline}\label{eta_trans}
    \dot{\eta}_{i}= \big[\mathbf{J}f(\mathbf{X}^{s})-\epsilon_{1}\lambda_{i}\mathbf{J}H(\mathbf{X}^s)\big]\eta_{i}\\
    -\epsilon_{2}\sum\limits_{j}\Tilde{\mathscr{L}}^{(2)}_{ij}(\mathbf{J}_1G(\mathbf{X}^s,\mathbf{X}^s)+\mathbf{J}_2G(\mathbf{X}^s,\mathbf{X}^s))\eta_{j}, \\
     i=2,3,\dots, MN.
\end{multline}
Here, $\eta_{1}$ represents the parallel mode, while the transverse modes are denoted by $\eta_{j}$ for $i = 2, 3, ..., MN$. Consequently, the stability analysis for the synchronized state is simplified to solving the coupled linear differential equation \eqref{eta_trans} associated with the transverse modes to determine the maximum Lyapunov exponent (MLE). This equation is referred to as the master stability equation. The stability criterion for the synchronization state mandates that the MLE transverse to the synchronous solution must be negative for stability to be established.
\par The coupled master stability equation \eqref{eta_trans} cannot be further simplified due to the following reasons: (i) the Laplacian matrices $\mathscr{L}^{(1)}$ and $\mathscr{L}^{(2)}$ generally do not commute, preventing simultaneous diagonalization with respect to a single eigenbasis, (ii) the Jacobians $\mathbf{J}H(\mathbf{X}^s)$ and $\mathbf{J}_1G(\mathbf{X}^s,\mathbf{X}^s)+\mathbf{J}_2G(\mathbf{X}^s,\mathbf{X}^s)$ are unrelated, making it impossible to combine the Laplacian matrices to construct an effective Laplacian matrix with orthonormal eigenbasis that could simplify the equation.
\par To simplify the stability analysis further, we proceed with the second scenario where the Jacobians $\mathbf{J}H(\mathbf{X}^s)$ and $\mathbf{J}_1G(\mathbf{X}^s,\mathbf{X}^s)+\mathbf{J}_2G(\mathbf{X}^s,\mathbf{X}^s)$ are related with each other. This condition can be achieved by selecting pairwise and three-body diffusive coupling functions in such a way that they satisfy an additional requirement, namely, $G(\mathbf{X},\mathbf{X})=H(\mathbf{X})$, which is commonly referred to as the natural coupling condition in the literature \cite{gambuzza2021stability,anwar2024global,muolo2023turing}. Now, if the coupling functions satisfy the natural coupling conditions, then we have
 \begin{equation}\label{natural_1}
 	\begin{array}{l}	\mathbf{J}_1G(\mathbf{X}^s,\mathbf{X}^s)+\mathbf{J}_2G(\mathbf{X}^s,\mathbf{X}^s)=\mathbf{J}H(\mathbf{X}^s)=\mathbf{J}_{s} \; \text{(say)}.
 	\end{array}
 \end{equation} Thus, in this scenario, we can start our analysis from the variational equation \eqref{equ5}. Substituting the above condition \eqref{natural_1} in Eq.~\eqref{equ5} yields, 
\begin{multline}  \delta\Dot{\mathbf{X}}_{i}=\mathbf{J}f(\mathbf{X}^{s})\delta\mathbf{X}_i-\sum\limits_{j=1}^{MN}(\epsilon_1\mathscr{L}^{(1)}_{ij}+\epsilon_2\mathscr{L}^{(2)}_{ij})\mathbf{J}_s\delta\mathbf{X}_j\\~~~~~~~~~~~~~~~~~~~~~~~~~~~~~~~~~~~~~~i=1,2,\cdots,MN.
\label{equ6}
\end{multline} 
This allows us to introduce an effective matrix $\mathbf{L}=(\epsilon_1\mathscr{L}^{(1)}+\epsilon_2\mathscr{L}^{(2)})$, which is again a real symmetric zero-row sum matrix, possessing the smallest eigenvalue $\gamma_{1}=0$. Furthermore, $\mathbf{L}$ is diagonalizable with respect to its orthonormal eigenbasis. Imposing the eigenbasis for $\mathbf{L}$ allows us to take an additional step, projecting the coupled linear system \label{eu6} onto fully decoupled linear systems with dimensions equal to the dimension of individual node dynamics. Each of these linear systems depends on a single eigenvalue, $\gamma_{i}$, of $\mathbf{L}$, more precisely:
\begin{equation}
\begin{array}{l}
\dot{\eta}_{i}=(\mathbf{J}f(\mathbf{X}^{s})-\gamma_i\mathbf{J}_s)\eta_i,~~~i=1,2,\dots,MN,
\label{decoupled_mse}      
\end{array}
\end{equation}
where $\eta_{i}=\sum\limits_{j}\delta\mathbf{X}_{j}v^{(i)}_{j}$ is the projection of $\delta\mathbf{X}_{j}$ on the eigenvector $v^{(i)}$ of the effective Laplacian matrix $\mathbf{L}$. This obtained master stability equation \eqref{decoupled_mse} is fully decoupled and thus exhibits a resemblance to the classical MSF approach \cite{pecora1998master}, except for the fact that here the eigenvalues of the Laplacian depend on the coupling strengths. Solving Eq.~\eqref{decoupled_mse} for maximum transverse Lyapunov exponent (which corresponds to $\eta_{i}$, $i=2,3,\dots, MN$) provides the necessary condition for the global synchronous solution to be stable. The negative values of the maximum transverse Lyapunov exponent indicate the emergence of the stable synchronous solution.   
\section{Results}\label{result}
Here, we present a series of numerical results that serve to validate our theoretical findings. We consider two different chaotic dynamical systems, namely the Hindmarsh-Rose (HR) model \cite{hr} and the R\"{o}ssler model \cite{rossler1976equation} as the individual dynamics of the nodes, coupled through pairwise and three-body interactions within and across the layers of a multilayer network. The HR-neurons are coupled by a linearly diffusive coupling scheme and the R\"{o}ssler oscillators are coupled by a non-linear diffusive coupling scheme. For the sake of simplicity, we proceed with only a two-layered $(M=2)$ system, as depicted in Fig.~\ref{fig1}. Global synchrony occurs when all the unitary components (nodes) in the multilayer network display identical oscillations over time, i.e.,
${\mathbf{X}}_{ij}(t) ={\mathbf{X}}^s(t)~(i=1,2;j = 1, 2,\dots, N).$ To quantify the global synchronization state, we introduce the instantaneous synchronization error $E(t)=\dfrac{1}{(2N-1)}\sum\limits_{i=1}^2\sum\limits_{j=1}^2\sum\limits_{k,l=1}^{N}\norm{\mathbf{X}_{ik}-\mathbf{X}_{jl}}$ which is zero when the multilayer network exhibits a globally synchronized state and nonzero finite for asynchronous dynamics. In the following, we will typically consider the time average of the synchronization error better to estimate the transition between the synchronous and asynchronous states. The multilayer network, given by Eq.~\eqref{equ1}, is therefore evaluated {using the fourth-order Runge-Kutta method} for a period of $3\times10^5$ time steps with integration step size $\delta t = 0.01$, and the last $10^5$ time units are taken for calculating the average synchronization error. We consider identical network topologies within each layer to delve into the influence of higher-order interactions, particularly three-way interactions and the parameter $\alpha$. The nodes within the layers are connected pairwisely to each other randomly with a probability $P_{1}$ following the algorithm of Erd\H{o}s R\'enyi random networks \cite{erdos1959random}. Similarly, the pairwise connections spanning across the layers are also established randomly, with a probability denoted as $P_{2}$. Three-body connections (2-simplices) are formed both within and across the layers by promoting all the triangles formed during the pairwise network construction process. Unless specified otherwise, the number of nodes in each layer is set to $N = 100$, with connection probabilities within the layers and across the layers being $P_{1}=0.1$ and $P_{2}=0.05$, respectively. In the case of the neuronal multilayer network, the pairwise and three-body coupling functions are considered as $H(\mathbf{X})=(m+n)\Gamma X$, and $G(\mathbf{X}_{i},\mathbf{X}_{j})=m\Gamma\mathbf{X}_{i}+n\Gamma\mathbf{X}_{j}$, where $\Gamma$ is the inner coupling matrix, characterizing through which variable the nodes interact with each other. Clearly, our chosen coupling functions satisfy the natural coupling condition. Without loss of generality, we consider $m=n=\frac{1}{2}$ and the inner coupling matrix is structured in such a way that all elements except $\Gamma_{11}$ are zero, with $\Gamma_{11}$ being equal to $1$. Thus, the nodes within and across the layers interact with one another through the first variable of the unitary node dynamics. The chosen linear diffusive coupling function resembles the electrical synaptic coupling for neuronal connections \cite{pereda2014electrical,bennett1997gap}. It is important to note that a linear higher-order interaction can be combined as pairwise interactions, eventually providing weighted pairwise interactions. Thus, choosing a linear higher-order interaction may not provide good insight into the effect of higher-order interactions. Here, we choose a linear higher-order coupling for neuronal networks mainly for two reasons. First, the linear diffusive coupling mimics the electrical synaptic coupling in neurons. Second, our developed analytical theory for dimension reduction of stability problems is limited to the case of diffusive coupling (linear or nonlinear). One can choose a nonlinear coupling such as chemical synaptic or noninvasive couplings. However, providing analytical support in those cases needs further restrictions such as regular topologies \cite{anwar2022stability}, which is beyond the scope of the present study. However, to show that both the linear and nonlinear couplings result in almost similar effects on the global synchronization, we illustrate numerical results with nonlinear noninvasive coupling for HR neuronal network in Appendix \ref{hr_nonlinear}.       
\par On the other hand, in the case of coupled R\"{o}ssler oscillators, we consider the pairwise and non-pairwise coupling functions to be nonlinear diffusive, defined as $h^{(1)}(\mathbf{X}_i,\mathbf{X}_j)=(0,y_j^3-y_i^3,0)^T$, and $h^{(2)}(\mathbf{X}_{i},\mathbf{X}_{j},\mathbf{X}_{k})=(0,y_j^2y_k-y_i^3,0)^T$, respectively. Note that this nonlinear coupling scheme also fulfills the aforementioned natural coupling condition.
\par In the subsequent analysis, we explore the emergence of global synchronization in the considered random multilayer structure. Additionally, we delve into the impact of various coupling parameters, including $\alpha, \epsilon_1,$ and $\epsilon_2$. Results pertaining to an alternative connection topology within the layers, specifically small-world connectivity \cite{watts1998collective}, are discussed in Appendix~\ref{small_world}. Apart from that, how the interlayer and intralayer synchronization emerge due to the combined effect of higher-order and multilayer structure has been discussed in Appendix \ref{intra_inter}.  
\subsection{Hindmarsh-Rose model with linear diffusive coupling scheme}\label{hr}
The Hindmarsh-Rose (HR) neuron model is a mathematical model used to describe the behavior of a simplified neuron. This model is known for its ability to exhibit complex neuronal dynamics, including periodic spiking, bursting, chaos, and other nonlinear behaviors. It is often used in computational neuroscience and mathematical biology to study neural activity. The three coupled differential equations that describe the evolution of the three variables ($x,y,z$ describe the membrane potential, slow current for recovery variable, and adapting current, respectively) of the model are given by
\begin{equation}
    \begin{array}{l}
         \frac{{dx}}{{dt}}=y-ax^3+bx^2-z+I,\\
         \frac{{dy}}{{dt}}=c-dx^2-y,\\
         \frac{{dz}}{{dt}}=r(s(x-x_0)-z)).
    \end{array}
\end{equation}
The model has eight parameters: $a,b,c,d,r,s,x_0,$ and $I$. Here, the parameter $I$ represents an external current input to the neuron. It can be used to simulate the effect of synaptic inputs or other external influences on the neuron's behavior. Other control parameters used often in the literature are $a,b,c,d,r$, the first four models of the working of the fast ion channels, and the last one of the slow ion channels, respectively. The control parameter $x_0$ delays or advances the activation of the slow current in the modeled neuron. These eight parameter values are taken as $a=1.0,b=3.0,c=1.0,d=5.0,r=0.005,s=4.0,x_0=-1.6,$ and $I=3.25$ to obtain chaotic behavior.
\begin{figure}[ht]
    \centering{
    \includegraphics[scale=0.25]{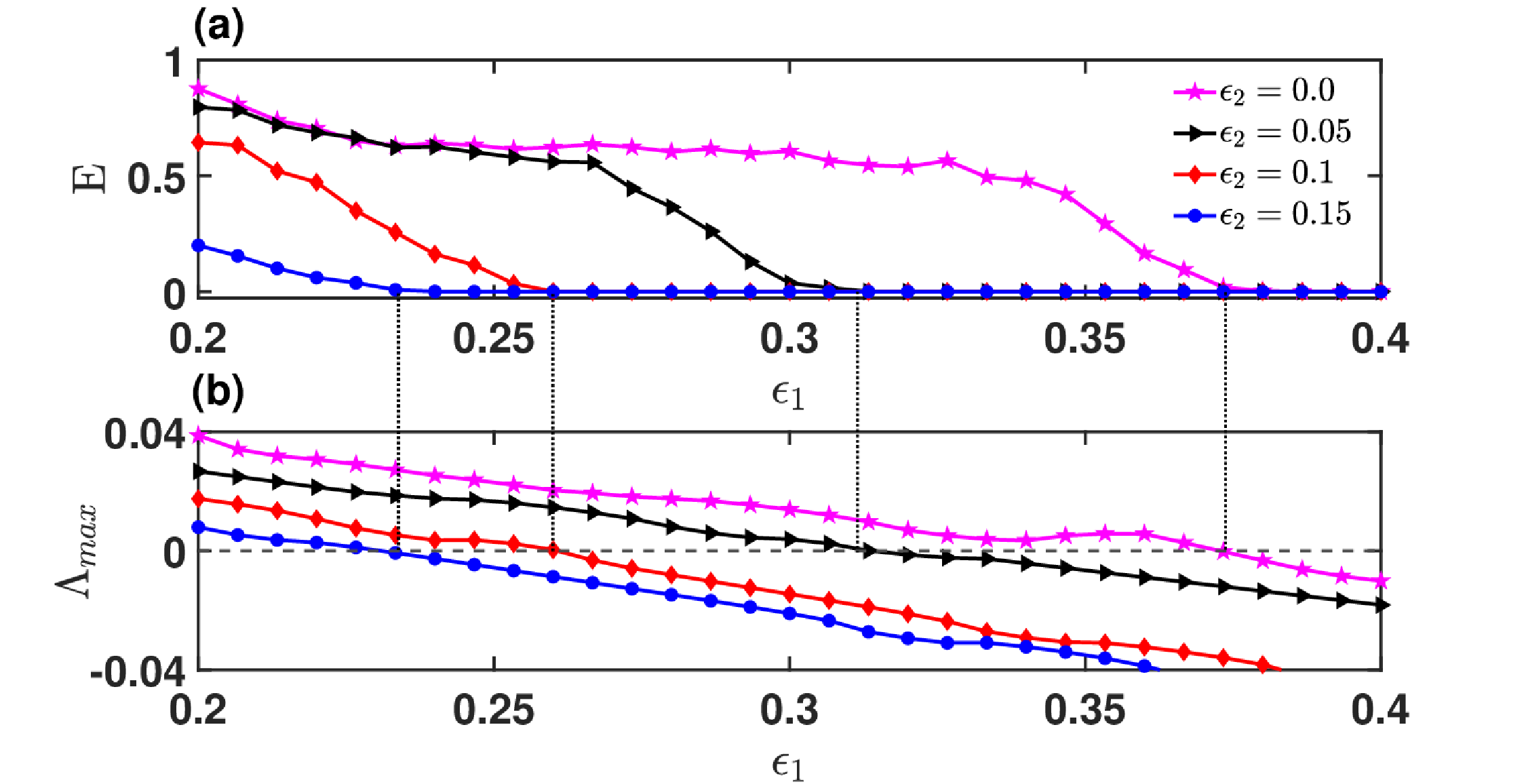}}
    \caption{The global synchronization of the HR neurons in the multilayer network for $N=100$ and $\alpha=0.3$. (a) global synchronization error (E) with respect to $\epsilon_1$ (b) maximum Lyapunov exponent ($\Lambda_{max}$) of the linearized Eq. \eqref{decoupled_mse} with respect to $\epsilon_1$. Four curves correspond to four values of $\epsilon_2$, $\epsilon_2=0.0$ (magenta curve), $\epsilon_2=0.05$ (black curve), $\epsilon_2=0.1$ (red curve), and $\epsilon_2=0.15$ (blue curve).}
    \label{fig2}
\end{figure}
\par  We consider that the dynamics of each of the nodes in the multilayer network \eqref{equ1} is governed by the HR neuronal model where neurons are coupled through the membrane potential by gap junctions, allowing for rapid and direct communication between neurons by facilitating the exchange of ions and small molecules. To study the emergence of global synchronization phenomena in the multilayer framework \eqref{equ1}, we start by evaluating the synchronization error $E$ by varying the strength of synaptic coupling of the pairwise $\epsilon_1$ between the neurons for different synaptic coupling strength of higher-order $\epsilon_2$ and a fixed value of $\alpha=0.3$. The corresponding results are depicted in Fig. \ref{fig2}(a). When there are no interactions involving three neurons (i.e., $\epsilon_2=0.0$), global synchronization is achieved at a critical value of pairwise synaptic coupling strength, $\epsilon_1=0.373$ (magenta curve in Fig. \ref{fig2}(a)). We then introduce the three-neuronal interactions and observe the effect of higher-order synaptic interactions on achieving global synchronization by increasing the three-body coupling strength, $\epsilon_2$. As $\epsilon_2$ increases (e.g., $\epsilon_2=0.05$), the synchronization state is achieved at a relatively smaller coupling strength $\epsilon_1=0.31$ (shown in a black curve). As $\epsilon_{2}$ is tuned up further, the system achieves synchronization much earlier, eventually leading to a significant enhancement in synchrony. For $\epsilon_2=0.1$ and $0.15$, the global synchronization state emerges at $\epsilon_1=0.26$ and $0.233$, depicted by red and blue curves, respectively. 
\par To confirm the validity of these findings, we assess the stability of the synchronous solution using the master stability function approach and analyze the maximum Lyapunov exponent ($\Lambda_{max}$) of the variational Eq. \eqref{decoupled_mse}, transverse to the synchronous manifold. In Figure \ref{fig2}(b), we present a plot of $\Lambda_{max}$ as a function of pairwise synaptic coupling strength ($\epsilon_1$), with $\epsilon_2$ values as mentioned earlier. The curves of $\Lambda_{max}$ attain negative values at the same critical $\epsilon_1$ values, where the synchronization error ($E$) becomes zero (indicated by dashed vertical lines). Thus, the analysis using the master stability function approach confirms our earlier observation that introducing three-neuronal interactions in the multilayer network advances the emergence of global synchrony.
\begin{figure}[ht]
    \centering{
    \includegraphics[scale=0.25]{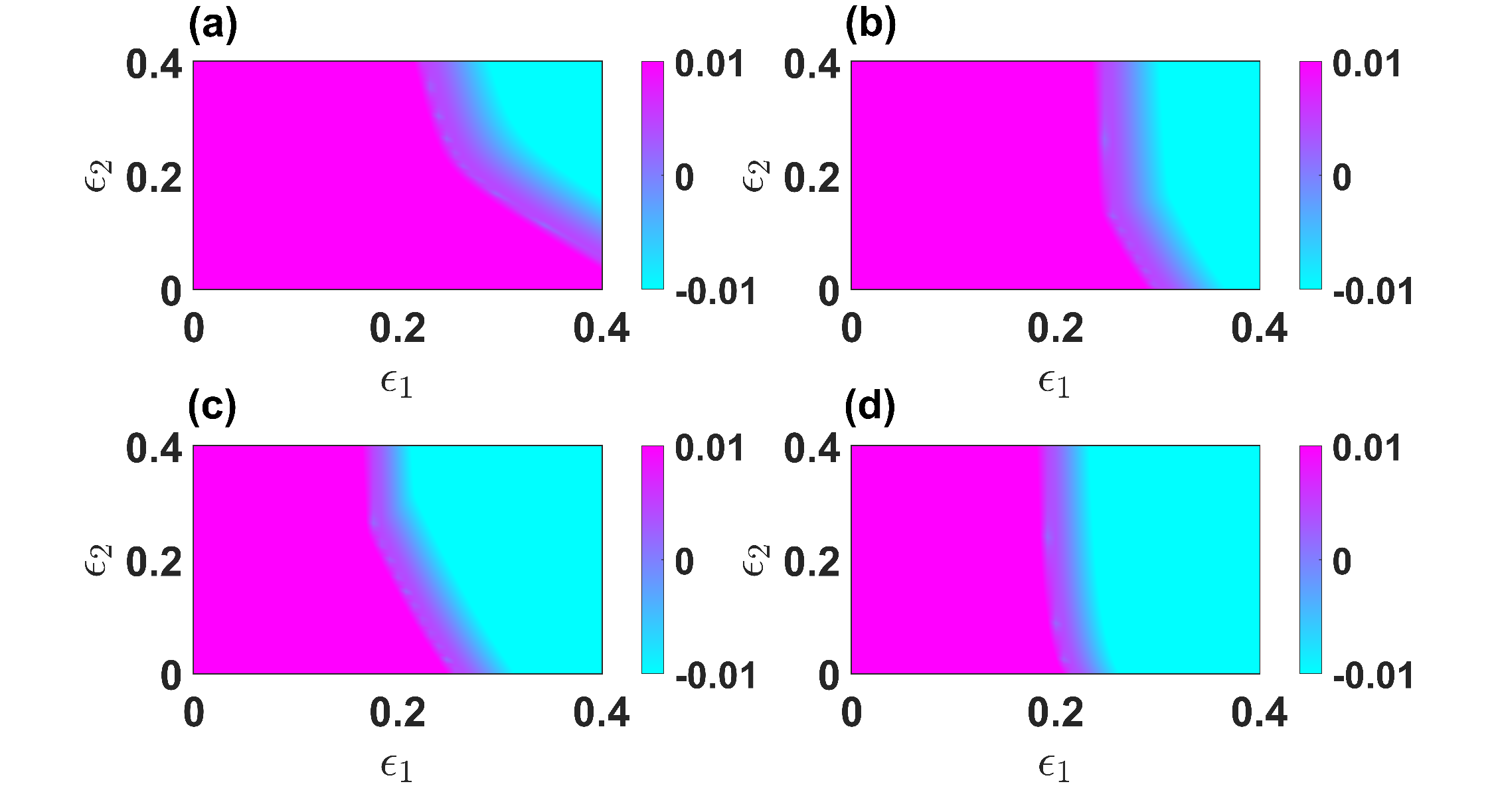}}
    \caption{The regions of the global synchronous and asynchronous states for $N=100$ HR-neurons in each layer of the multilayer network \eqref{equ1}. The maximum Lyapunov exponent $\Lambda_{max}(\epsilon_1,\epsilon_2,\alpha)$ of the linearized Eq. \eqref{decoupled_mse} in the $(\epsilon_1,\epsilon_2)$ parameter space for four different values of $\alpha$: (a) $\alpha=0.2$, (b) $\alpha=0.4$, (c) $\alpha=0.5$, and (d) $\alpha=0.9$.}
    \label{fig3}
\end{figure}
\par To study the combined effect of pairwise and higher-order synaptic coupling on the emergence of the global synchronization state in the multilayer network,  we compute the maximum Lyapunov exponent $\Lambda_{max}$ (obtained from Eq.~\eqref{decoupled_mse}) as a function of $\epsilon_{1}$ and $\epsilon_{2}$ for four different values of the parameter $\alpha$. The corresponding results are represented in Fig.~\ref{fig3}, where the variation of $\Lambda_{max}$ is delineated through the color bars. Our findings reveal that within a certain range of $\epsilon_1$, the higher-order coupling strength $\epsilon_2$ does not impact the synchronization state for all values of $\alpha \in (0,1)$. However, a significant influence of $\epsilon_2$ becomes evident beyond this range. Specifically, the neurons in the network do not synchronize for any value of $\epsilon_2$ when the pairwise synaptic coupling strength falls within a certain range, which depends on the specific value of $\alpha$. As this range is exceeded, the critical value of $\epsilon_2$, or the necessary higher-order synaptic coupling strength for synchrony, gradually decreases to zero as $\epsilon_1$ increases. We also observed a similar impact of both synaptic coupling strengths ($\epsilon_1,\epsilon_2$) on synchrony by varying the value of $\alpha$. The main difference observed for the four different values of $\alpha$ lies in the emerged region of global synchrony (cyan-colored region) or asynchrony (magenta-colored region). We consider $\alpha=0.2$ in Fig.~\ref{fig3}(a) and observe that after a certain range $(0<\epsilon_1<0.25)$ of $\epsilon_1$ at the beginning, the required value of $\epsilon_2$ to get the global synchrony is found to be decreasing as $\epsilon_1$ increases up to $\epsilon_1=0.4$. In \ref{fig3}(b), for $\alpha=0.4$, within the range $0<\epsilon_1<0.25$, there is no effect of the group interaction between the neurons on the emergence of synchrony. Beyond $\epsilon_{1}=0.25$, with increasing $\epsilon_{2}$, the global synchrony emerges at a relatively lower critical pairwise coupling. Furthermore, it is observable that beyond $\epsilon_1=0.3$ the critical value of $\epsilon_2$ to achieve the global synchrony is decreased to zero, i.e., beyond $\epsilon_1=0.3$, the global synchronization can emerge even if there are no three-body interactions. In \ref{fig3}(c), and \ref{fig3}(d), we take $\alpha=0.5,0.9$, respectively, and the critical values of $\epsilon_1$ up to which the higher-order synaptic coupling strength, $\epsilon_2$ remain inactive, are found to be at $\epsilon_1=0.19,0.18$, respectively. Also, the critical values on the $\epsilon_1$-axis, where the multilayer network is in the global synchronization state without group interaction, are found at $\epsilon_1=0.26,0.22$, respectively. Notably, the synchrony region increases while the desynchrony region decreases as $\alpha$ ranges from $0$ to $1$. Thus, as the strength of interlayer coupling increases, the multilayer network can achieve global synchronization more easily. 
\begin{figure}[ht]
    \centering{
    \includegraphics[scale=0.24]{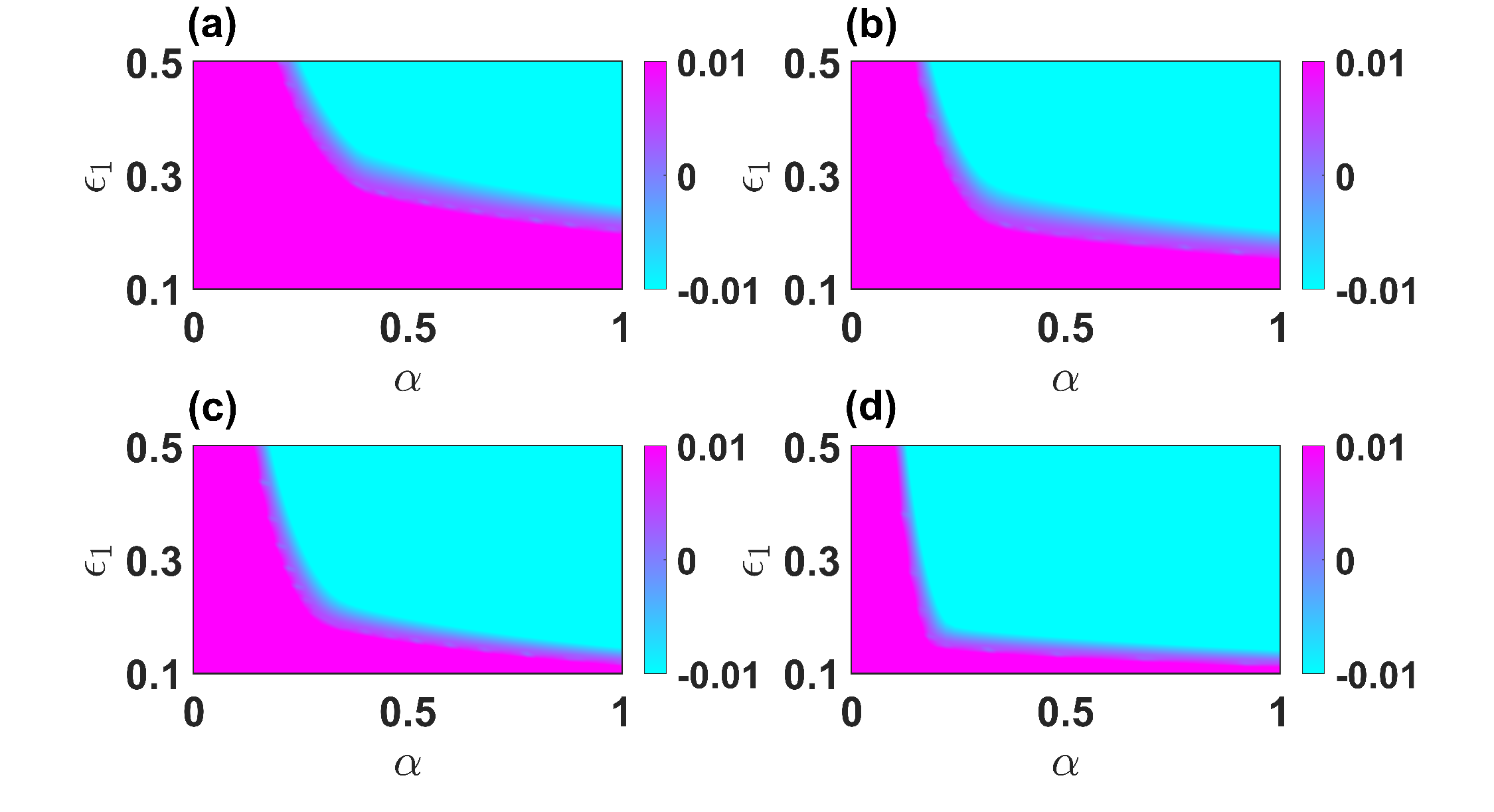}}
\caption{The global synchrony (stable) and asynchrony (unstable) region in $(\alpha,\epsilon_1)$-plane for $N=100$ HR-neurons in each layer of the multilayer network \eqref{equ1}. The maximum Lyapunov exponent $\Lambda_{max}(\epsilon_1,\epsilon_2,\alpha)$ of the linearized Eq. \eqref{decoupled_mse} in $(\alpha,\epsilon_1)$ parameter space for four values of $\epsilon_2$: (a) $\epsilon_2=0.0$, (b) $\epsilon_2=0.1$, (c) $\epsilon_2=0.2$, and (d) $\epsilon_2=0.4$.}
    \label{fig4}
\end{figure}
\par Next, we study the combined effect of $\alpha$ and $\epsilon_1$ on the global synchronization for different values of $\epsilon_2$. The variation of $\Lambda_{max}$ of the linearized Eq. \eqref{decoupled_mse} in the $(\alpha,\epsilon_1)$-parameter space are shown for four values of $\epsilon_2$ in Fig. \ref{fig4}. In Fig. \ref{fig4}(a), we consider the case when the strength of the higher-order synaptic coupling is zero ($\epsilon_2=0$), which means when there is no effect of higher-order neuronal interaction in the multilayer model. It is observable that as the value of the parameter $\alpha$ (or, $\epsilon_1$) increases, the critical value of the pairwise coupling strength $\epsilon_1$ ( $\alpha$) to achieve global synchronization decreases. Also when three-neuron interactions are introduced, i.e., for non-zero values of $\epsilon_{2}$ in the remaining subfigures of Fig. \ref{fig4}, one can observe a similar kind of effect of the parameters $\alpha,\epsilon_1$ on the global synchronization state and notably an enhancement of the stable region of global synchronization is noticeable at higher values of $\epsilon_2$. In Fig. \ref{fig4}(b), we set $\epsilon_2=0.1$. Here, for a fixed value of the parameter $\alpha$ (or, $\epsilon_1$), the critical value of $\epsilon_1$ (or, $\alpha$) decreases with increasing higher-order coupling strength. In Fig. \ref{fig4}(c) and Fig. \ref{fig4}(d), we increase higher-order coupling strength to $\epsilon_2=0.2$, and $\epsilon_2=0.4$, respectively, and a qualitatively similar behavior is observed. This observation strengthens our statement that introducing three-neuron interactions amplifies the likelihood of synchronization among the neurons within a multilayer neuronal network.
\par
\begin{figure}[ht]
    \centering{
    \includegraphics[scale=0.25]{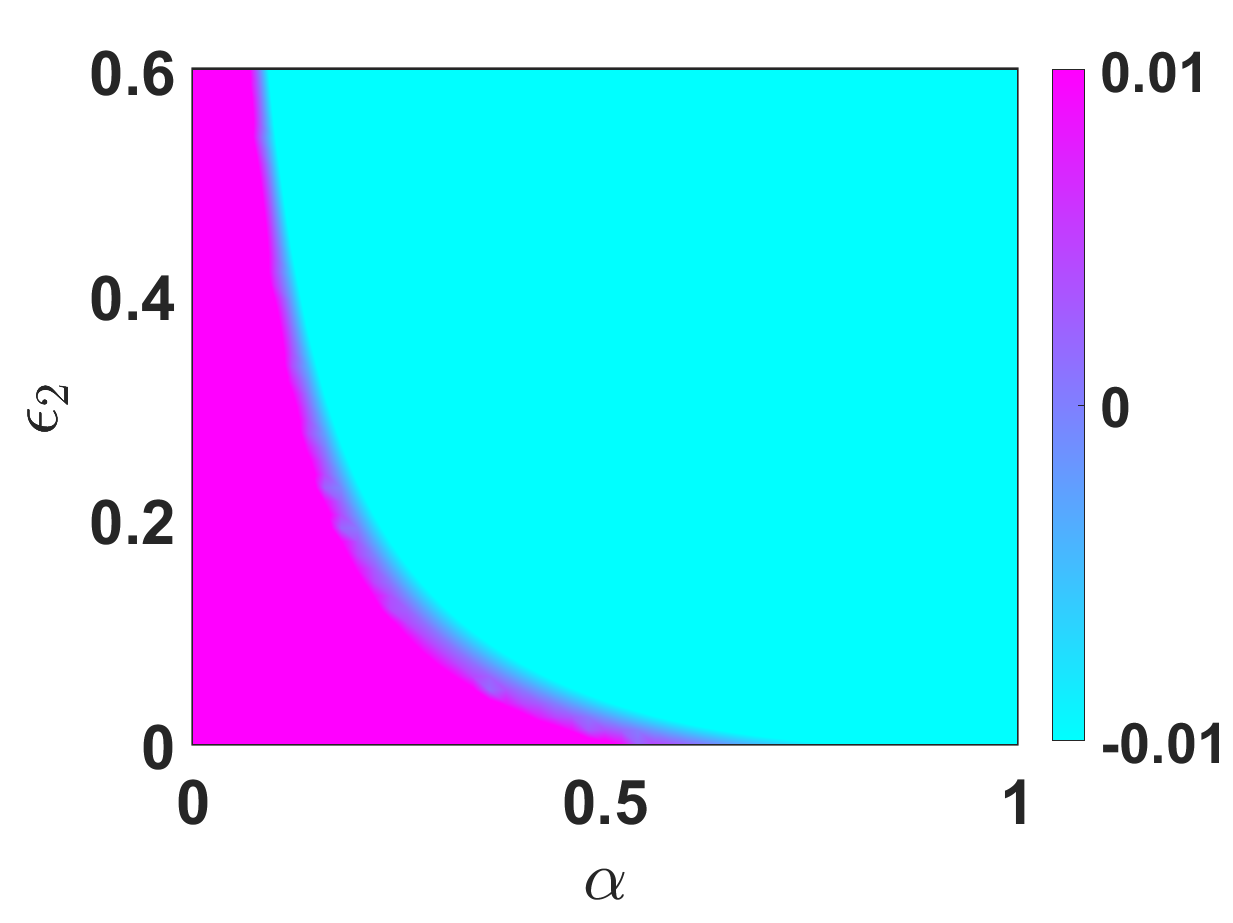}}
    \caption{The variation of the maximum Lyapunov exponent, $\Lambda_{max}(\epsilon_1,\epsilon_2,\alpha)$ of the linearized Eq. \eqref{decoupled_mse} in $(\alpha,\epsilon_2)$ parameter space for a fixed $\epsilon_1=0.2$, with $N=100$ HR-neurons in each layer of the multilayer network \eqref{equ1}.}
    \label{fig5}
\end{figure}

 The combined effects of the parameters $\alpha$ and $\epsilon_2$ for a fixed value of the pairwise synaptic coupling strength ($\epsilon_1=0.2$) are shown in Fig. \ref{fig5}. One can observe that global synchronization is forbidden for $0<\alpha<0.06$, and beyond $\alpha=0.06$, the required strength of the higher-order synaptic coupling for global synchronization decreases with the increase in $\alpha$. When $\alpha$ is tuned up further and crosses $\alpha=0.6$, the multilayer network achieves global synchrony even without higher-order interactions (i.e., $\epsilon_{2}=0$).
\subsection{R\"{o}ssler model with non-linear diffusive coupling scheme}\label{rossler}
This section aims to show that the previously presented results hold true beyond the example of the dynamical system shown above, i.e., the HR neuronal model. For the sake of definiteness, we thus used the paradigmatic chaotic R\"{o}ssler system as the underlying dynamics governing individual nodes within the multilayer network. The dynamics of the R\"{o}ssler model are governed by the following equations,
\begin{equation}
    \begin{array}{l}
         \frac{{dx}}{{dt}} = -y - z, \\
         \frac{{dy}}{{dt}} = x + ay, \\
         \frac{{dz}}{{dt}} = b + z(x - c),
    \end{array}
\end{equation}
where for $a=0.2$, $b=0.2$ and $c=5.7$, the system exhibits chaotic behavior.
In our previous analysis using the HR model, we assumed the underlying pairwise and higher-order interactions to be linear diffusion as it mimics the synaptic interactions among the neurons. However, in the context of the R\"{o}ssler system, we explore nonlinear pairwise and higher-order coupling schemes to delve into the impact of nonlinear interactions on the emergence of global synchrony within multilayer networks. Specifically, the coupling functions $h^{(1)}(X_i, X_j)$ and $h^{(2)}(X_i, X_j, X_k)$ are assumed to be of the functional forms \( (0,y_j^3-y_i^3,0)^T \) and \( (0,y^2_jy_k-y^3_i,0)^T \), respectively so that it satisfies the natural coupling condition.
\begin{figure}[hpt]
	\centering{
		\includegraphics[scale=0.18]{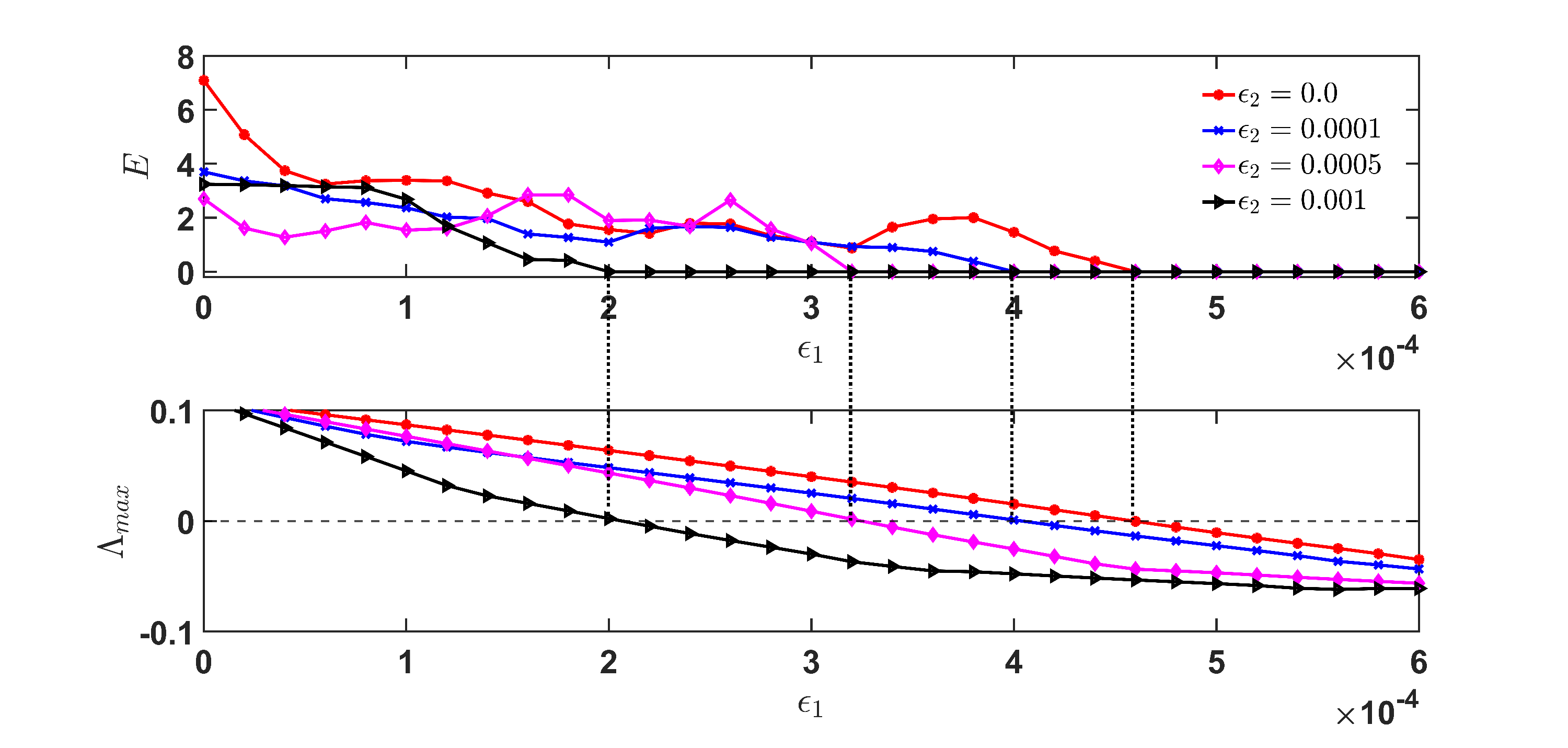}}
	\caption{The global synchronization of the multilayer network for $N=100$ R\"{o}ssler oscillators and $\alpha=0.3$. (a) global synchronization error (E) with respect to $\epsilon_1$ (b) maximum Lyapunov exponent ($\Lambda_{max}$) of the linearized Eq. \eqref{decoupled_mse} with respect to $\epsilon_1$. Four curves correspond to four values of $\epsilon_2$, $\epsilon_2=0.0$ (red curve), $\epsilon_2=0.05$ (blue curve), $\epsilon_2=0.1$ (magenta curve), and $\epsilon_2=0.15$ (black curve).}
	\label{fig6}
\end{figure}
\par In a manner akin to the previous exploration of the HR multilayer network, the investigation initiates by assessing global synchronization error ($E$) while varying the coupling strength of pairwise interactions ($\epsilon_1$) for various levels of coupling strength of higher-order interactions ($\epsilon_2$), with a constant $\alpha$ value of $0.3$. The obtained results are illustrated in Fig. \ref{fig6}(a). When higher-order interactions are absent (i.e., $\epsilon_2=0.0$), global synchronization is attained at a critical value of pairwise coupling strength, $\epsilon_1=0.00046$ (red curve in Fig. \ref{fig6}(a)). Subsequently, the introduction of higher-order interactions reveals the impact of higher-order interactions on achieving global synchronization by increasing the higher-order coupling strength ($\epsilon_2$). As $\epsilon_2$ tuned up from zero, the synchronization state is achieved at relatively lower values of $\epsilon_1$. For $\epsilon_2=0.0001$, synchronization emerges at $\epsilon_1=0.0004$, which is depicted by the blue curve. The magenta curve corresponds to $\epsilon_2=0.0005$ where global synchronization occurs at $\epsilon_1=0.00032$, and the black curve represents $\epsilon_2=0.001$ with the critical $\epsilon_1$ value found to be $\epsilon_1=0.0002$. To validate these findings, the stability of the synchronous solution is assessed using the master stability function approach, and the maximum Lyapunov exponent ($\Lambda_{max}$) of the variational Eq. \eqref{decoupled_mse} transverse to the synchronous manifold is analyzed. Figure \ref{fig6}(b) presents $\Lambda_{max}$ plotted against pairwise coupling strength ($\epsilon_1$), with varying $\epsilon_2$ values as aforementioned. The curves of $\Lambda_{max}$ exhibit negative values at the same critical $\epsilon_1$ values, coinciding with zero synchronization error ($E$) (marked by dotted vertical lines). The validation of the master stability function approach confirms the earlier observation that introducing higher-order interactions in the multilayer network advances the emergence of global synchrony, underscoring the significant impact of changing higher-order coupling strengths on the global synchronization of the multilayer network.
\begin{figure}[hpt]
	\centering{
		\includegraphics[scale=0.25]{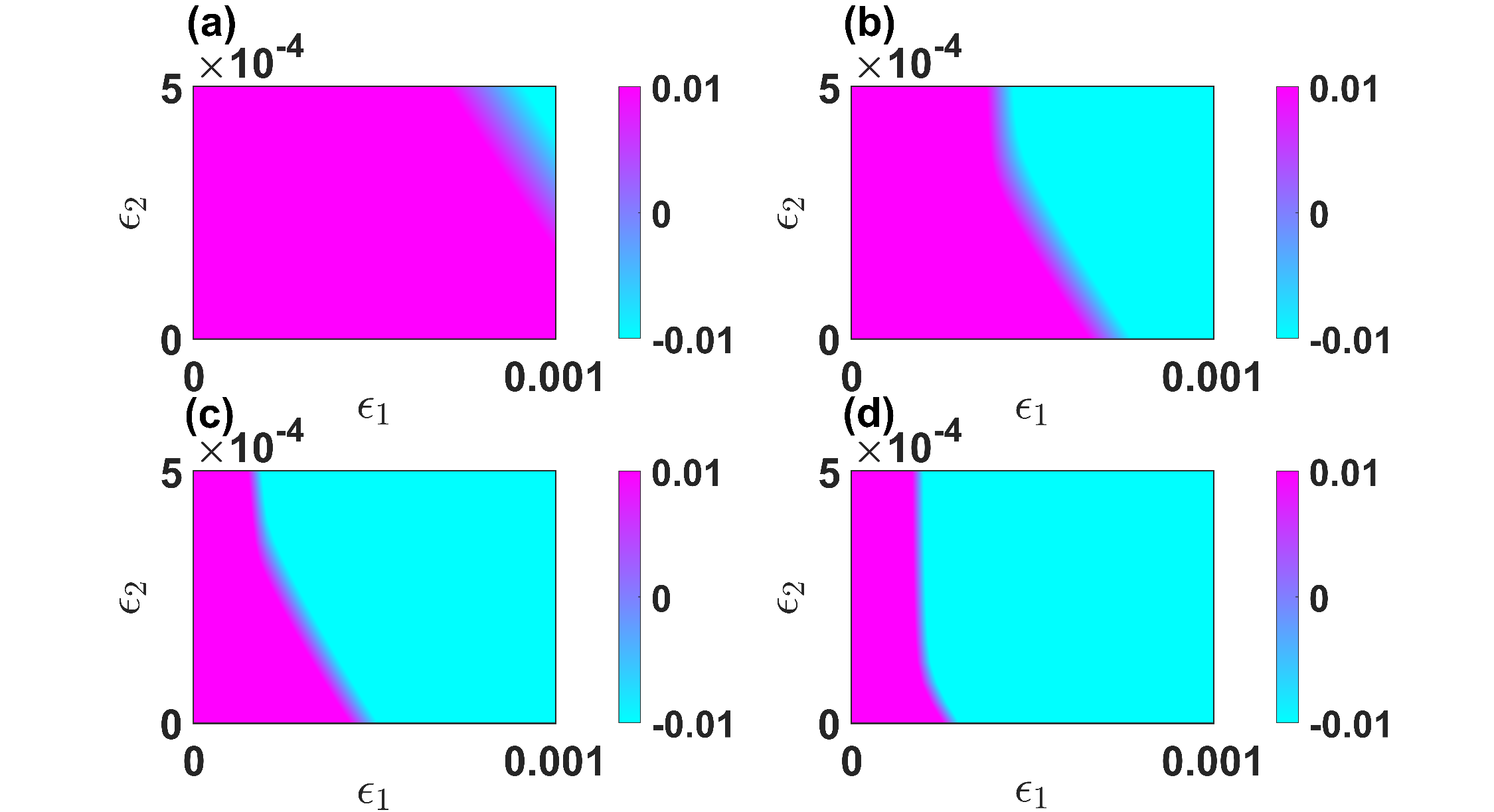}}
	\caption{The stable and unstable region of the global synchronization state in $(\epsilon_1,\epsilon_2)$ parameter space for $N=100$ R\"{o}ssler oscillators in both the layers of the network \eqref{equ1} with the random intralayer and interlayer network topologies. The maximum Lyapunov exponent $\Lambda_{max}$ of the linearized Eq. \eqref{decoupled_mse} in the $(\epsilon_1,\epsilon_2)$-parameter space are for four values of $\alpha$: (a) $\alpha=0.1$, (b) $\alpha=0.2$, (c) $\alpha=0.3$, and (d) $\alpha=0.6$.}
	\label{fig7}
\end{figure}
\par Then, in the parameter space (\(\epsilon_1,\epsilon_2\)), we investigate the influence of both pairwise and higher-order coupling strengths on the global synchronization state for four different values of \(\alpha\). Using color bars in Fig. \ref{fig7}, we depict variations in the maximum Lyapunov exponents \(\Lambda_{max}\), which is obtained from the linearized Eq. \eqref{decoupled_mse}. Our findings reveal that within a certain range of \(\epsilon_1\), the higher-order coupling strength \(\epsilon_2\) does not affect the synchronization state for all \(\alpha\) values within the interval \((0,1)\) as in the previous case of HR multilayer network. However, a noticeable influence of \(\epsilon_2\) emerges beyond this range. Specifically, nodes (or, R\"{o}ssler oscillators) in the network fail to synchronize for any \(\epsilon_2\) value when \(\epsilon_1\) falls within a specific range, dependent on the precise \(\alpha\) value. As this range is surpassed, the critical \(\epsilon_2\) value, or the requisite higher-order coupling strength for synchrony, gradually diminishes to zero as \(\epsilon_1\) increases. We also observe a similar impact of both coupling strengths (\(\epsilon_1,\epsilon_2\)) on synchrony by varying \(\alpha\). It is observable that the region of synchronization expands while the desynchrony region contracts as \(\alpha\) ranges from \(0\) to \(1\), revealing that stronger interlayer connections play an important role in the emergence of global synchronization.
\begin{figure}[hpt]
	\centering{
		\includegraphics[scale=0.19]{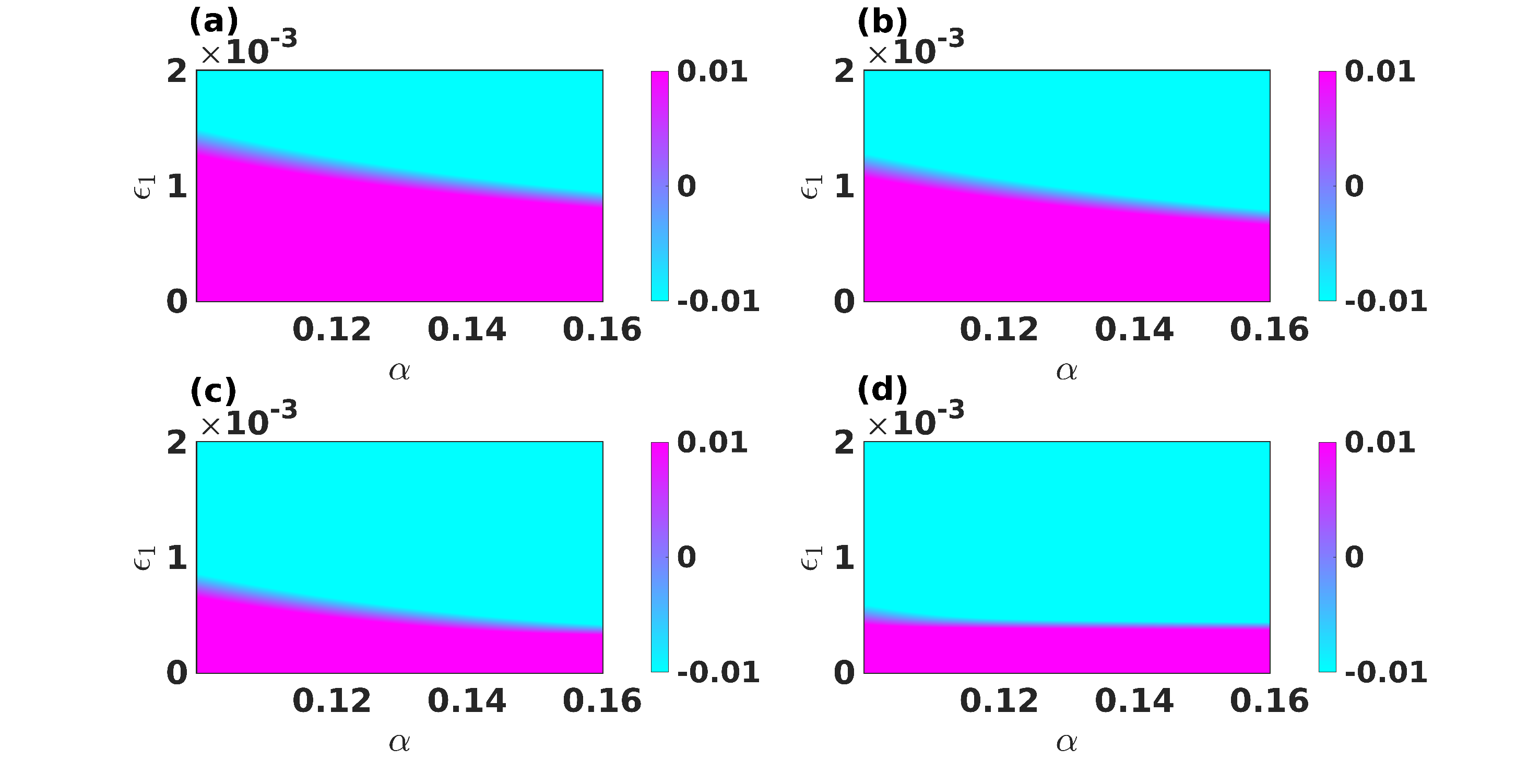}}
	\caption{The regions of the global synchronous and asynchronous states for $N=100$  R\"{o}ssler oscillators in each layer of the multilayer network \eqref{equ1}. The maximum Lyapunov exponent $\Lambda_{max}(\epsilon_1,\epsilon_2,\alpha)$ of the linearized Eq. \eqref{decoupled_mse} in the $(\alpha,\epsilon_1)$ parameter space for four different values of $\epsilon_2$: {(a) $\epsilon_2=0.0$, (b) $\epsilon_2=0.0001$, (c) $\epsilon_2=0.0005$, and (d) $\epsilon_2=0.001$.}}
	\label{fig8}
\end{figure}
\par Thereafter, we investigate the combined impact of \(\alpha\) and \(\epsilon_1\) on global synchronization for various values of \(\epsilon_2\). The variation of \(\Lambda_{max}\) in the \((\alpha,\epsilon_1)\)-parameter space is depicted for four distinct \(\epsilon_2\) values in Fig. \ref{fig8}. In Fig. \ref{fig8}(a), we examine the scenario where the strength of higher-order coupling is zero (\(\epsilon_2=0\)), indicating no influence of higher-order interaction in the multilayer model. As \(\alpha\) (or \(\epsilon_1\)) increases, we observe a decrease in the critical value of pairwise coupling strength \(\epsilon_1\) (or \(\alpha\)) required for global synchronization. When higher-order interaction is introduced, i.e., for non-zero values of higher-order coupling strength in the remaining subfigures of Fig. \ref{fig8}, a similar trend is observed for the parameters \(\alpha\) and \(\epsilon_1\) regarding the global synchronization state. Notably, we find an expansion of the stable region of global synchronization at higher values of \(\epsilon_2\). This observation once again reinforces our assertion that introducing higher-order interactions enhances the likelihood of synchronization among nodes within multilayer networks.
\par
\begin{figure}[ht]
    \centering{
    \includegraphics[scale=0.25]{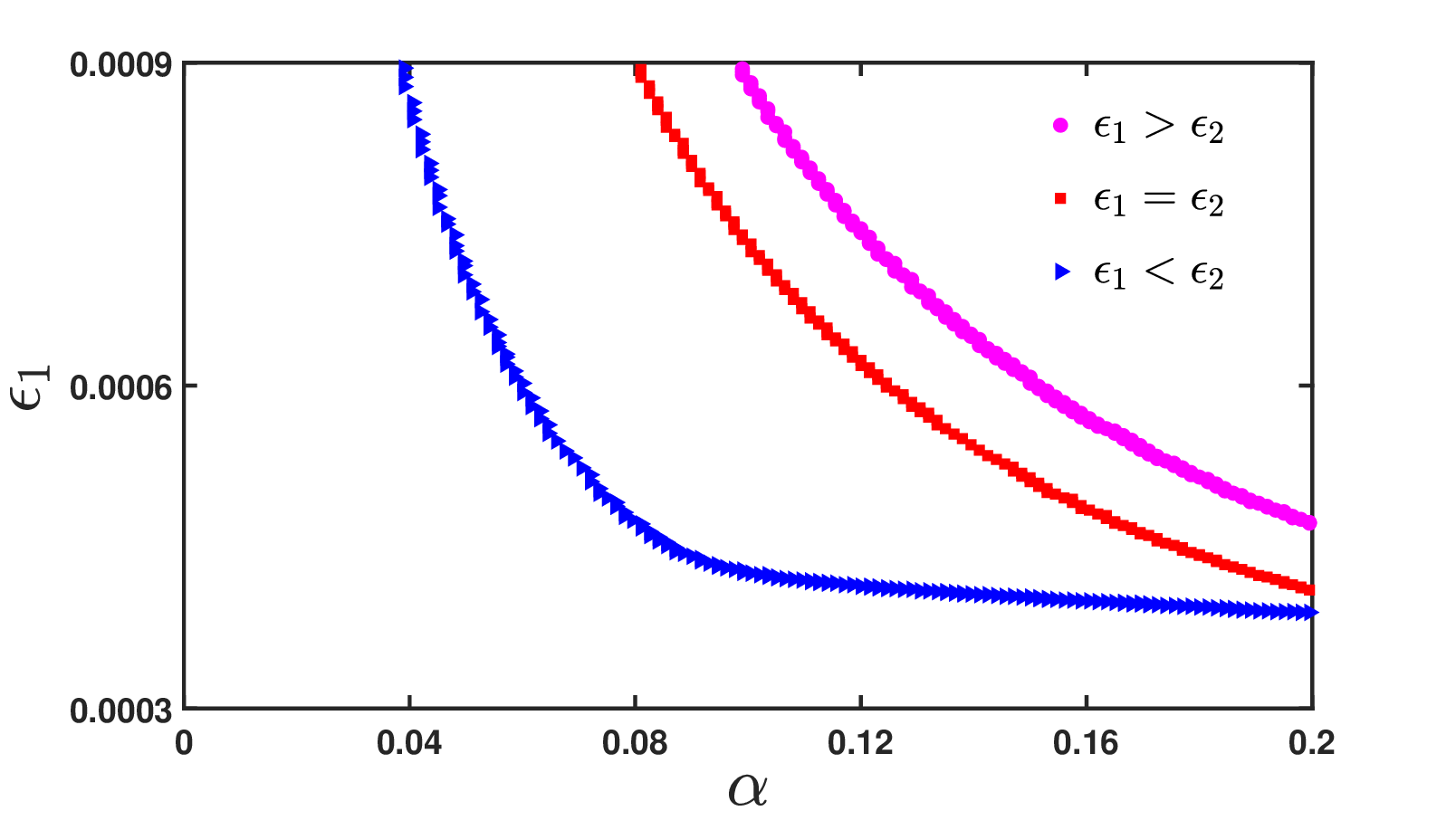}}
    \caption{{{\bf The threshold coupling values in the ($\alpha,\epsilon_1$)-parameter space for three cases: $\epsilon_1<\epsilon_2$, $\epsilon_1=\epsilon_2$, and $\epsilon_1>\epsilon_2$.} The magenta circular points represent the case when $\epsilon_1>\epsilon_2$ ($\epsilon_1=3.0\epsilon_2$), red square-shaped points are for the case $\epsilon_1=\epsilon_2$, and the blue triangular-shaped points are depicting the case when $\epsilon_1<\epsilon_2$ ($\epsilon_1=0.5\epsilon_2$).}}
    \label{fig9}
\end{figure}

 {To understand the effect of pairwise and higher-order interactions better, in Fig.~\ref{fig9}, we plot the synchronization threshold values in $(\alpha,\epsilon_{1})$ parameter plane by considering three different choices of higher-order coupling strength $\epsilon_{2}$, namely $\epsilon_{2}<\epsilon_{1}$, $\epsilon_{2}=\epsilon_{1}$ and $\epsilon_{2}>\epsilon_{1}$. The threshold couplings are the values of $(\alpha,\epsilon_{1})$ for which the maximum Lyapunov exponent $(\Lambda_{max})$ becomes zero while crossing from positive to negative regime. Thus, the region to the left (right) of the threshold curve indicates the region of desynchrony (synchrony). When $\epsilon_2<\epsilon_1$, we consider the relation $\epsilon_1=3.0\epsilon_2$, and plotted the threshold points in $(\alpha,\epsilon_1)$-plane by magenta circular points. The red square shaped points are the threshold points when $\epsilon_1=\epsilon_2$, and the blue diamond shaped points are the threshold points when $\epsilon_1<\epsilon_2$ satisfying $\epsilon_1=0.5\epsilon_2$. It is observed that the synchronous region in the $(\alpha,\epsilon_1)$-plane is maximum when $\epsilon_2>\epsilon_1$, i.e., when the higher-order interactions are more effective than the pairwise ones. On the other hand, when the pairwise interactions are more effective, the obtained global synchronization region is the smallest compared to the other two cases.} 
\par
\begin{figure}[hpt]
    \centering{
    \includegraphics[scale=0.35]{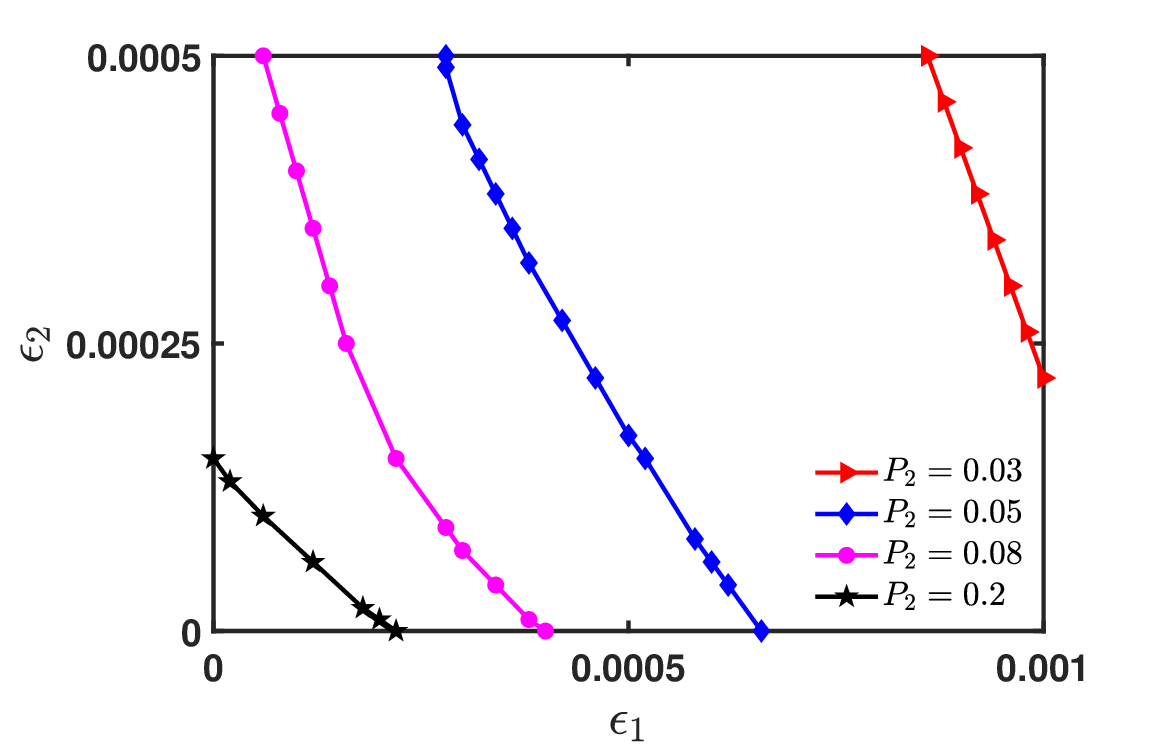}}
    \caption{{{\bf The threshold coupling values in the $(\epsilon_1,\epsilon_2)$-parameter space for four distinct random interlayer topologies of the multilayer network.} Each layer of the multilayer network consists of \(N=100\) R\"{o}ssler oscillators, with a random topology for both intralayer and interlayer connections. The probability for the random intralayer connections is fixed at $P_1=0.1$. The points on the red (triangle) curve represent the threshold points when the probability for the random interlayer connections is $P_2=0.03$, the points on the blue (diamond) curve are for $P_2=0.05$, the magenta (square) one is for $P_2=0.08$, and the threshold points on the black (star) curve are for $P_2=0.2$. The interlayer coupling strength is set fixed at $\alpha=0.2$.}}
    \label{fig10}
\end{figure}

 { So far, we have discussed the results associated with a fixed interlayer connection probability, $P_2$, meaning that the number of links and triangles joining the two layers remains constant. However, $P_{2}$ is a crucial parameter as it controls the number of links and triangles between the layers.  When $P_{2}=0$, the layers are disconnected from one another, while for $P_{2}=1$, each node in a layer is connected to all the nodes in the other layer. Therefore, with increasing $P_{2}$, the number of pairwise and higher-order interactions increases between the layers. Thus, to examine the effect of $P_{2}$ on the emergence of global synchronization, we calculate the critical value of coupling strengths $(\epsilon_{1},\epsilon_{2})$ for different values of $P_{2}$, specifically, $P_2=0.03,0.05,0.08,$ and $0.2$. Here, the probability for the random intralayer connection is fixed at $P_1=0.1$, i.e., the intralayer topology remains invariant. The corresponding critical curves are portrayed in Fig.~\ref{fig10} for fixed interlayer coupling strength $\alpha=0.2$. The regions to the left of these critical curves represent the domain of desynchronization, while the regions to the right correspond to the domain of global synchronization. It is evident that with increasing $P_{2}$, the region of global synchrony becomes wider. This implies that a higher number of pairwise and higher-order connections between the layers makes it easier to achieve global synchronization. Thus, the global synchronous region can be expanded by introducing more links and triangles in the interlayer connection topology. Furthermore, one can observe an interesting result for $P_2=0.2>P_1$, i.e., when the number of pairwise and higher-order interactions between the layers is higher than within the layers. In this scenario, the global synchronization emerges even if $\epsilon_1=0.0$ (black curve). This implies that the presence of sufficient triangles in the multilayer network alone can promote global synchronization, even when there are no pairwise interactions.}

 \section{Discussion}\label{sec5}
In summary, we introduce a generalized mathematical model aimed at capturing intricate higher-order interactions within and across layers of a multilayer framework. Our investigation focuses on global synchronization. Within the constraints of the invariance condition for the synchronous solution, we derive the necessary criteria for achieving a stable synchronization state, thereby extending the well-established master stability function approach to multilayer structures with higher-order interactions. The complexity introduced by multiple layers and higher-order interactions in our examined system is exemplified by the manifestation of the master stability formalism as a set of coupled linear differential equations rather than a singular parametric variational equation. Despite this intricacy, we demonstrate that, in a specific scenario, our formalism simplifies to yield a set of uncoupled parametric variational equations, each possessing dimensions equivalent to those of a single dynamical unit. In addition to the theoretical foundations, we incorporate a collection of numerical results that affirm the validity and applicability of our methodology. Our findings demonstrate that introducing higher-order interactions within and across layers of a multilayer network expands the parameter space wherein global synchronization can be achieved.
\par
Therefore, our in-depth analytical exploration has provided valuable insights into the impact of higher-order interactions on the emergence of global synchronization within a generalized multilayer structure. However, it is important to acknowledge the possibility of numerous unexplored avenues for further investigation. Within this context, a natural extension of our framework involves examining synchronization phenomena in generic multilayer structures featuring distinct intralayer and interlayer coupling schemes. Additionally, delving into the influence of higher-order interactions in generalized multilayer networks on the emergence of other synchronization phenomena, such as intralayer and interlayer synchronization in detail, presents itself as a potentially intriguing research direction.

\appendix

\section{{Hindmarsh-Rose model with nonlinear coupling scheme}}\label{hr_nonlinear}
\begin{figure}[ht]
    \centering{
    \includegraphics[scale=0.35]{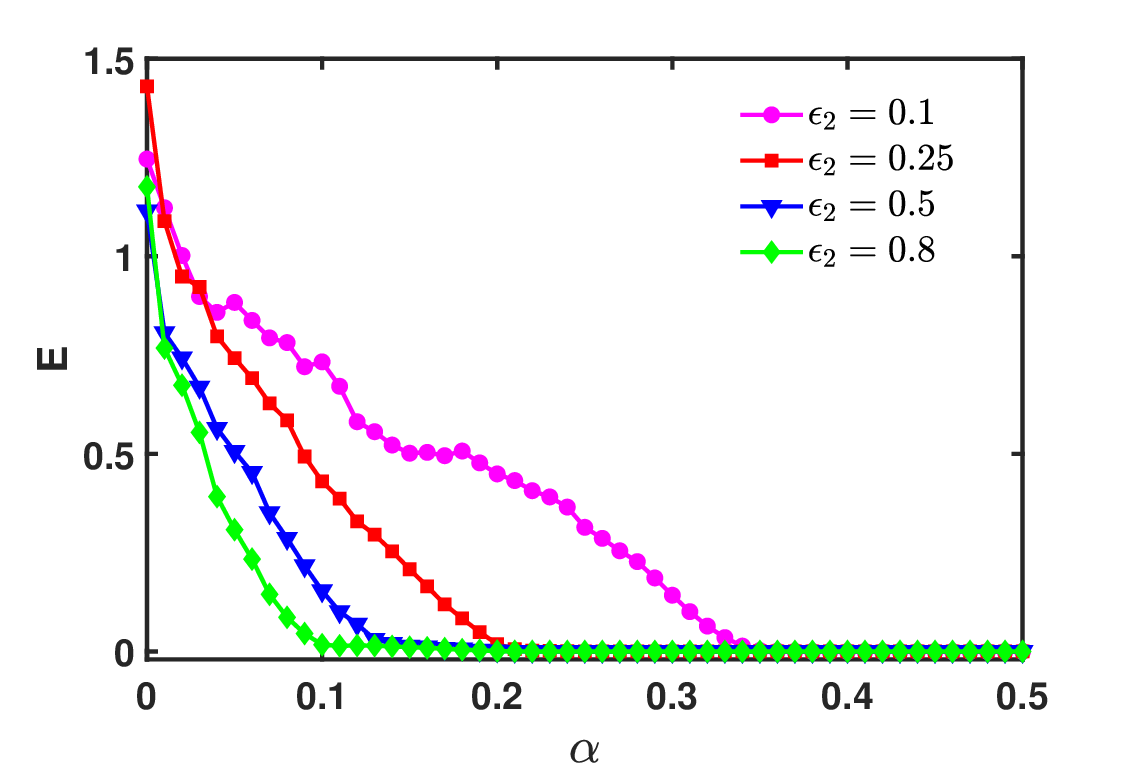}}
    \caption{{{\bf Variation of global synchronization error as a function of $\alpha$ for Hindmarsh-Rose neuronal model}. Each layer of the multilayer network consists of \(N=100\) HR-neurons, with a random topology for both intralayer and interlayer connections and nonlinear coupling functions. The pairwise coupling strength is kept fixed at $\epsilon_1=0.1$. The four curves colored in magenta (circle), red (square), blue (inverted triangle), and green (diamond) represent global synchronization error for four distinct values of the non-pairwise coupling strengths given by $\epsilon_2=0.1,0.25,0.5,$ and $0.8$, respectively.}}
    \label{fig11}
\end{figure}

{In our proposed multilayer HR-neuronal model given in Section \ref{hr}, both pairwise and non-pairwise coupling functions were initially considered linear due to our analytical reliance on diffusive couplings and resemblance with the gap junction (electrical synaptic) couplings. However, in this section, we adopt a nonlinear, non-diffusive coupling scheme previously used by Gambuzza et al. \cite{gambuzza2021stability}. The coupling functions are defined as $h^{(1)}(X_i, X_j)=\left(\tanh \left(\frac{{x}_{j}-{x}_{i}}{0.5}\right),0,0\right)$, $h^{(2)}(X_i, X_j, X_k)=\left(\tanh \left(\frac{{x}_{j}+{x}_{k}-2{x}_{i}}{0.5}\right),0,0\right)$. This coupling is noninvasive, so it guarantees the emergence of global synchronization without any further assumptions. We, therefore, investigate the impact of non-pairwise coupling strength on the emergence of global synchronization error in the multilayer network with this coupling scheme. The result corresponding to this scenario is presented in Fig. \ref{fig11}, where we plot global synchronization error $(E)$ with respect to the parameter $\alpha$ by varying the higher-order coupling strength $\epsilon_2$. Fixing the pairwise coupling strength at $\epsilon_1=0.1$, we  consider four different values of the non-pairwise coupling strengths, namely, $\epsilon_2=0.1,0.25,0.5,$ and $0.8$. One can observe that the global synchronous state emerges at relatively smaller values of the interlayer strength $\alpha$ as the non-pairwise coupling strength $\epsilon_{2}$ is increased. Thus, analogous to the linear higher-order coupling, the nonlinear higher-order couplings within and across the layers make achieving the global synchronization state easier.}

\section{Small-world connectivity} \label{small_world}
\begin{figure}[ht]
    \centering{
    \includegraphics[scale=0.25]{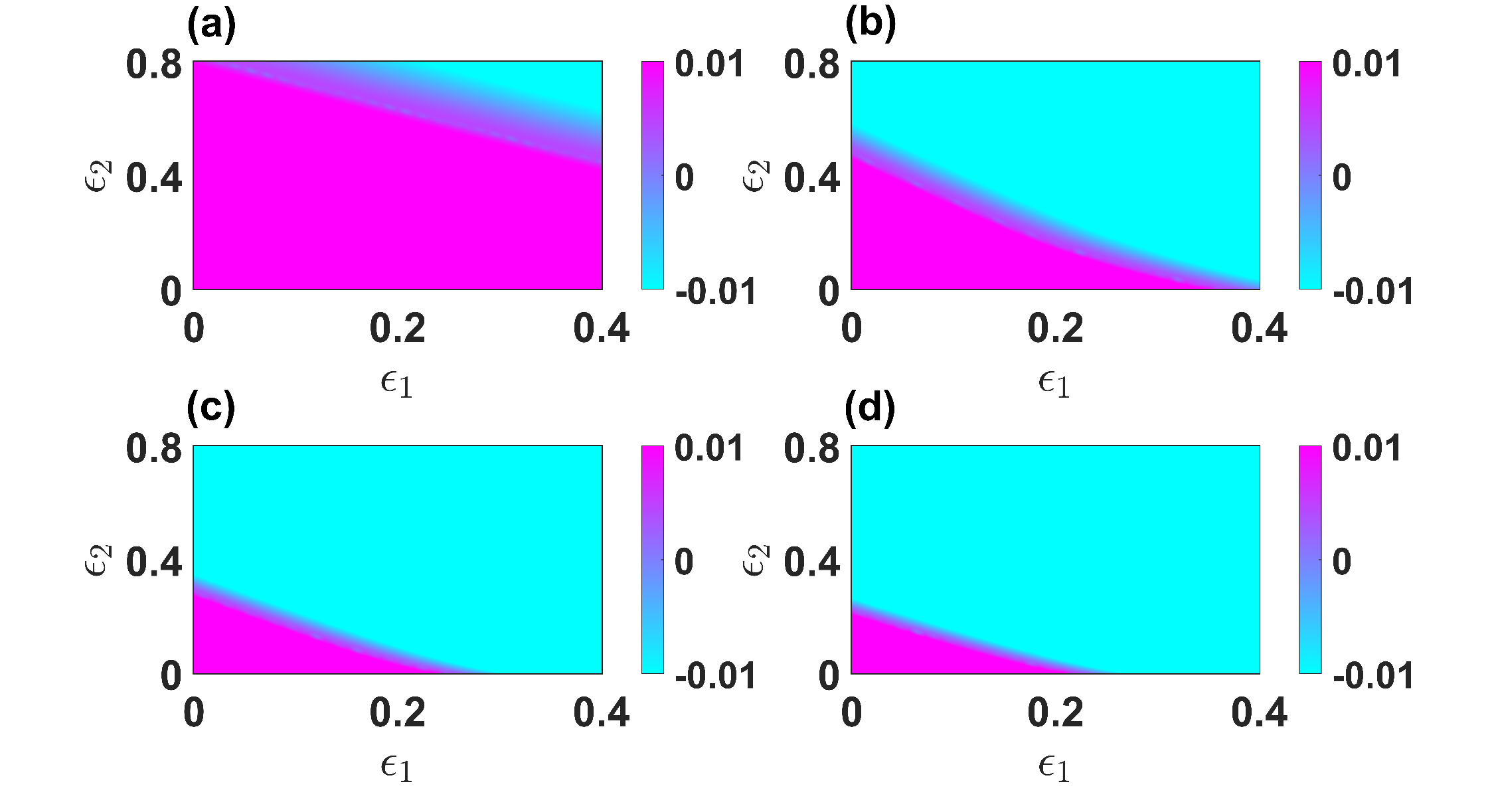}}
    \caption{The stable and unstable region of the global synchronization state in $(\epsilon_1,\epsilon_2)$ parameter space for $N=100$ HR neurons in both the layers of the network \eqref{equ1} with the intralayer network topology is taken as Watts-Strogatz (WS) model for small-world networks. The maximum Lyapunov exponent $\Lambda_{max}$ of the linearized Eq. \eqref{decoupled_mse} in the $(\epsilon_1,\epsilon_2)$-parameter space are for four values of $\alpha$: (a) $\alpha=0.1$, (b) $\alpha=0.3$, (c) $\alpha=0.6$, and (d) $\alpha=0.9$.}
    \label{fig12}
\end{figure}

Here, we broaden the scope of our investigation by examining results under an alternative connectivity topology among nodes within the layers of the multilayer structure. Specifically, we depart from random connections among nodes and, instead, adopt the Watts-Strogatz small-world algorithm \cite{watts1998collective} to establish connections to the nodes within each layer of the multilayer framework. Nevertheless, the connections across the layers remain randomized, consistent with our previous approach. The aim is to demonstrate that the findings we have previously obtained are not confined solely to the random network structures among nodes within the layer. Small-world connectivity within each layer is formed so that each node can interact with its $10$ $(k=10)$ nearest neighbors. Additionally, a probability parameter, $p_{sw} = 0.1$, allows nodes to connect with a few more distantly located within the layers. We assume the HR model governs the dynamics of each node. Additionally, we consider the same pairwise and higher-order coupling schemes as earlier. Considering the small-world intralayer topology in each layer and taking the HR model as the dynamics of each neuron, the maximum Lyapunov exponent of the Eq. \eqref{decoupled_mse} has been depicted in Fig. \ref{fig12}. We take $\alpha=0.1, 0.3,0.6$, and $0.9$ in Fig. \ref{fig12}(a), Fig. \ref{fig12}(b), Fig. \ref{fig12}(c), and Fig. \ref{fig12}(d), respectively. Combining these four figures, we can conclude that the qualitative combined effect of the two synaptic coupling strengths $\epsilon_1$ and $\epsilon_2$ is the same as the previous (Fig. \ref{fig3}), i.e., the region of synchronization increases with increasing higher-order coupling. However, a significant difference can be observed. Here, the enhancement rate is fast from the previous, and in this case, the impact of the higher-order synaptic coupling strength is observable even in the absence of pairwise interactions, i.e., the sole presence of higher-order interactions is enough for the emergence of global synchronization. The effect of the parameter $\alpha$ on the synchronous region in the $(\epsilon_1,\epsilon_2$)-plane is completely the same as the previous case; a rise in $\alpha$ value results in an enhancement of the synchronous region.

\section{{Intralayer and interlayer synchronization states}} \label{intra_inter}
\begin{figure}[ht]
    \centering{
    \includegraphics[scale=0.35]{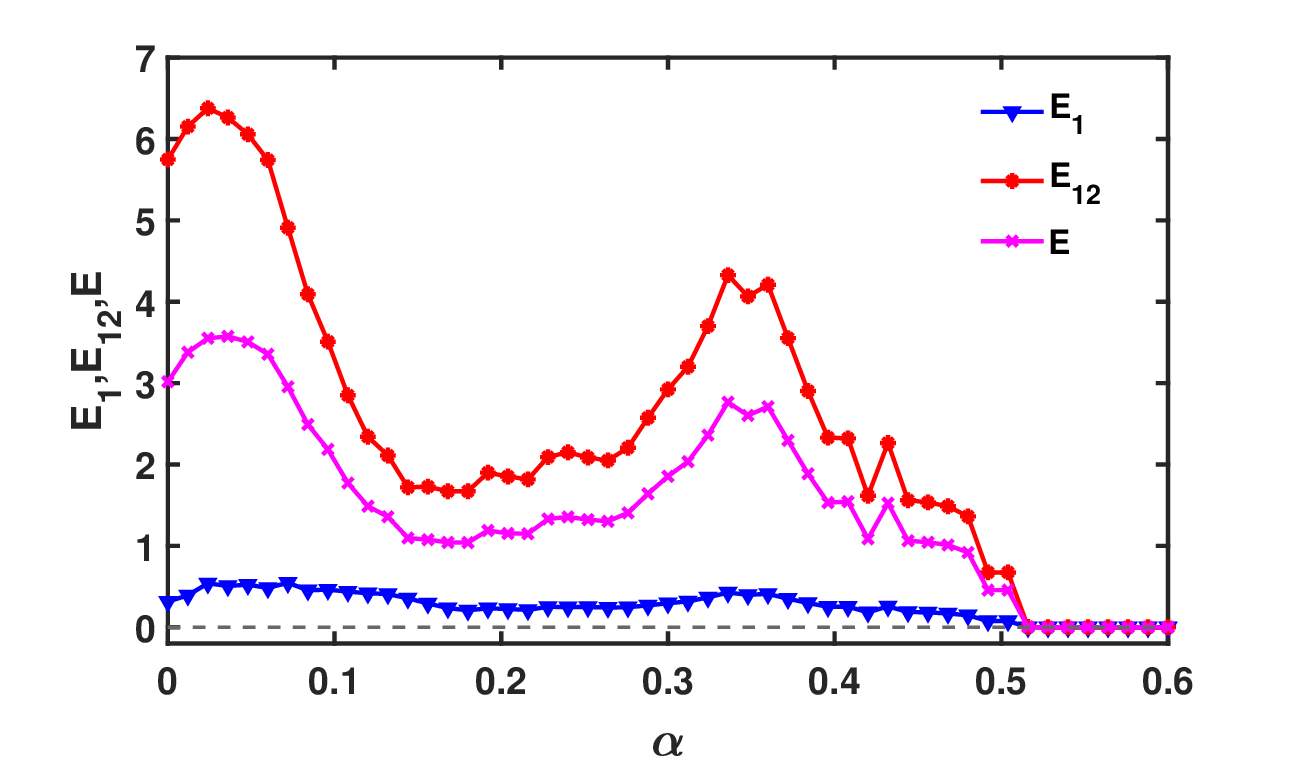}}
    \caption{{{\bf The intralayer, interlayer, and global synchronization errors as a function of $\alpha$.} The multilayer is consisting of  $N=100$ R\"{o}ssler oscillators in both the layers of the network \eqref{equ1} with the random intralayer and interlayer network topology. Pairwise and non-pairwise coupling strengths are fixed at $\epsilon_1=0.0003$ and $\epsilon_2=0.0001$, respectively. Three curves of colored blue (inverted triangle), red (circle), and magenta (star), correspond to the intralayer, interlayer, and global synchronization errors.}}
    \label{fig13}
\end{figure}

{Throughout the main text, all the results discussed focus on the emergence of global synchronization in the higher-order multilayer framework. Here, we delve shortly into the emergence of intralayer and interlayer synchronization in our considered multilayer network given by Eq.~\eqref{equ1}. To do so, we consider the previously taken multilayer network constituting $100$ R\"{o}ssler oscillators in each layer with nonlinear diffusive couplings given in the Sec. \ref{rossler}. To characterize the intralayer and interlayer synchronization states, we introduce two synchronization errors as follows, 
\begin{equation}
    \begin{array}{l}
        E_1(t)=\dfrac{1}{(N-1)}\sum\limits_{k,j=1}^{N}\norm{\mathbf{X}_{1k}-\mathbf{X}_{1j}}, \\
        E_{12}(t)=\dfrac{1}{N}\sum\limits_{k,j=1}^{N}\norm{\mathbf{X}_{1k}-\mathbf{X}_{2j}}. 
    \end{array}
\end{equation}
Here $E_{1}$ denotes the synchronization error corresponding to layer-$1$. Thus, when $E_{1}$ becomes zero, the nodes within the layer-$1$ become synchronized. Now, it is well known that in the multilayer network, each layer synchronizes simultaneously \cite{rakshit2019enhancing}. Hence, whenever layer-$1$ synchronizes, so does the layer-$2$? Eventually, the zero value of $E_{1}$ indicates the emergence of intralayer synchronization. On the other hand, $E_{12}$ represents the synchronization error between the layers, i.e., the interlayer synchronization error. Therefore, zero value $E_{12}$ indicates the emergence of interlayer synchronization in the multilayer network. Therefore, we investigate the variation of intralayer and interlayer synchronization errors by varying the system parameters to observe the occurrence of the synchronized states. In Fig.~\ref{fig13}, we  plot the intralayer synchronization error ($E_1$), the interlayer synchronization error ($E_{12}$) and global synchronization error ($E$) with respect to the interlayer coupling parameter $\alpha$. We fix the pairwise and non-pairwise coupling strength at $\epsilon_1=0.0003$ and $\epsilon_2=0.0001$. We choose these two strengths very small enough so that both layers are not in the synchronous state when $\alpha=0.0$, i.e., when the layers are totally disconnected from one another. Now, after gradually increasing the value of the parameter $\alpha$, it is observable that the interlayer and intralayer synchronization states, and simultaneously the global synchronization state emerge at the same value of $\alpha$ ($\alpha \approx 0.5$). Therefore, for fixed pairwise and higher-order coupling (drawn from the desynchrony region), inter and intralayer synchrony occur at the same threshold value of the connecting/disconnecting parameter $\alpha$ between the layers.}

 \section*{Acknowledgements}
 P.K. Pal sincerely thanks the University Grants Commission (UGC), India, for their generous and invaluable financial support, made available under the reference number NTA Ref. No. 191620041800. This support has been instrumental in facilitating and advancing the research efforts and contributions made during this study.
 
\bibliographystyle{apsrev4-2} 
\bibliography{multilayer}

\end{document}